\documentclass[fleqn,usenatbib]{mnras}

\usepackage{newtxtext,newtxmath}
\usepackage{booktabs}
\newcommand{\oxfordastro}{Department of Astrophysics, University of Oxford, Denys Wilkinson Building, Keble Road, Oxford OX1 3RH, UK}
\newcommand{\oxfordmag}{Magdalen College, University of Oxford, Oxford OX1 4AU, UK}
\newcommand{\birmingham}{School of Physics \& Astronomy, University of Birmingham, Edgbaston, Birmingham B15 2TT, United Kingdom}

\newcommand{\liege}{Astrobiology Research Unit, Université de Liège, 19C Allée du 6 Août, 4000 Liège, Belgium}
\newcommand{\miteaps}{Department of Earth, Atmospheric and Planetary Sciences, Massachusetts Institute of Technology, 77 Massachusetts Avenue, Cambridge,\\ MA 02139, USA }
\newcommand{\mitkavli}{Kavli Institute for Astrophysics and Space Research, Massachusetts Institute of Technology, Cambridge, MA 02139, USA }
\newcommand{\sandiego}{Department of Astronomy \& Astrophysics, UC San Diego, 9500 Gilman Drive, La Jolla, CA 92093, USA}
\newcommand{\iac}{Instituto de Astrof\'{i}sica de Canarias (IAC), Calle Vía Láctea s/n, 38200, La Laguna, Tenerife, Spain}
\newcommand{\cambridge}{Cavendish Laboratory, JJ Thomson Avenue, Cambridge CB3 0HE, UK}

\newcommand{\cfa}{Center for Astrophysics \textbar  Harvard \& Smithsonian, 60 Garden Street, Cambridge, MA 02138, USA}
\newcommand{\liegestar}{Space Sciences, Technologies and Astrophysics Research (STAR) Institute, Universit\'e de Li\'ege, All\'ee du 6 Ao\^ut 19C, B-4000 Li\'ege, Belgium}

\newcommand{\geneva}{Observatoire de Gen\`eve, D\'epartement d’Astronomie, Universit\'e de Gen\`eve, Chemin Pegasi 51b, 1290 Versoix, Switzerland}
\newcommand{\nasaames}{NASA Ames Research Center, Moffett Field, CA 94035, USA}

\newcommand{\grenoble}{Univ. Grenoble Alpes, CNRS, IPAG, F-38000 Grenoble, France}
\newcommand{\cadi}{Cadi Ayyad University, Oukaimeden Observatory, High Energy Physics, Astrophysics and Geoscience Laboratory, FSSM, Morocco}
\newcommand{\sharjah}{Department of Applied Physics and Astronomy, and Sharjah Academy for Astronomy, Space Sciences and Technology, University of Sharjah,\\ United Arab Emirates}

\newcommand{\ull}{Deptartamento de Astrof\'{i}sica, Universidad de La Laguna (ULL), 38206 La Laguna, Tenerife, Spain}
\newcommand{\mex}{Universidad Nacional Aut\'onoma de M\'exico, Instituto de Astronom\'ia, AP 70-264, Ciudad de M\'exico, 04510, M\'exico}

\newcommand{\bern}{Center for Space and Habitability, University of Bern, Gesellschaftsstrasse 6, 3012, Bern, Switzerland}

\newcommand{\aim}{AIM, CEA, CNRS, Universit\'e Paris-Saclay, Universit\'e de Paris, F-91191
Gif-sur-Yvette, France}

\newcommand{\iaa}{Instituto de Astrofísica de Andalucía (IAA-CSIC), Glorieta de la Astronomía s/n, 18008 Granada, Spain}

\newcommand{\sfasu}{Department of Physics, Engineering and Astronomy, Stephen F. Austin State University, 1936 North St, Nacogdoches, TX 75962, USA}
\newcommand{\pest}{Perth Exoplanet Survey Telescope, Perth, Western Australia}

\newcommand{\oac}{Universidad Nacional de C\'ordoba - Observatorio Astron\'{o}mico de C\'{o}rdoba, Laprida 854, X5000BGR, C\'ordoba, Argentina}
\newcommand{\conicet}{Consejo Nacional de Investigaciones Científicas y Técnicas (CONICET), Godoy Cruz 2290, CPC 1425FQB, CABA, Argentina}
\newcommand{\vandy}{Department of Physics and Astronomy, Vanderbilt University, Nashville, TN 37235, USA}
\newcommand{\icn}{Universidad Nacional Autónoma de México, Instituto de Ciencias Nucleares, Cto. Exterior S/N, C.U., Coyoacán, 04510 Ciudad de Mexico, México}
\newcommand{\uniturkey}{Department of Astronomy \& Space Sciences, Faculty of Science, Ankara University, TR-06100, Ankara, T\"urkiye}
\newcommand{\obsturkey}{Ankara University, Astronomy and Space Sciences Research and Application Center (Kreiken Observatory), Incek Blvd., TR-06837, Ahlatlıbel, Ankara, T\"urkiye}
\newcommand{\trottier}{Institut Trottier de recherche sur les exoplan\`etes, D\'epartement de Physique, Universit\'e de Montr\'eal, Montr\'eal, Qu\'ebec, Canada}

\newcommand{\riogrande}{Departamento de F\'isica Te\'orica e Experimental, Universidade Federal do Rio Grande do Norte, Campus Universit\'ario, Natal, RN, 59072-970, Brazil}
\newcommand{\instiporto}{ Instituto de Astrof\'isica e Ci\^ncias do Espa\c{c}o, Universidade do Porto, CAUP, Rua das Estrelas, 4150-762 Porto, Portugal}
\newcommand{\uniporto}{Departamento de F\'isica e Astronomia, Faculdade de Ci\^encias, Universidade do Porto, Rua do Campo Alegre, 4169-007 Porto, Portugal}
\newcommand{\uname}{Universidad Nacional Aut\'onoma de M\'exico, Instituto de Astronom\'ia, AP 106, Ensenada 22800, BC, M\'exico}
\newcommand{\mariecurie}{Paris Region Fellow, Marie Sklodowska-Curie Action}

\usepackage[T1]{fontenc}

\DeclareRobustCommand{\VAN}[3]{#2}
\let\VANthebibliography\thebibliography
\def\thebibliography{\DeclareRobustCommand{\VAN}[3]{##3}\VANthebibliography}

\usepackage{graphicx}	
\usepackage{amsmath}	

\title[TOI-1080~b]{TOI-1080~b: a temperate, rocky planet orbiting a quiet M4V host}

\author[Y.\ G\'{o}mez Maqueo Chew \& G.\ Dransfield et al.]{Y. G\'{o}mez Maqueo Chew$^{1}$\thanks{email: ygmc@astro.unam.mx}\color{blue}${\ddagger}$\color{black}, 
G. Dransfield$^{2,3,4}$\thanks{email: george.dransfield@physics.ox.ac.uk}\thanks{These two authors contributed equally to this work and should be considered joint first authors.},
K. Barkaoui$^{5,6,7}$,
C. Cadieux$^{8,9}$,
E. Ducrot$^{10,11}$,
\newauthor
B. V. Rackham$^{7,12}$,
M. Timmermans$^{4,6}$,
A. J. Burgasser$^{13}$,
A. Segura$^{14}$,
K. G. Stassun$^{15}$,
C. Ziegler$^{16}$,
\newauthor
A. Soubkiou$^{6}$,
J. M. Almenara$^{17,8}$,
B.O. Demory$^{18}$,
M. Gillon$^{6}$,
J. M. Jenkins$^{19}$,
E. Jofr\'e$^{20,21}$,
\newauthor
A. Khandelwal$^{1}$,
S. P\'aez$^{1}$,
R. Petrucci$^{20,21}$,
L. Parc$^{8}$,
M. Pichardo Marcano$^{1}$,
I. Plauchu-Frayn$^{22}$,
\newauthor
U. Schroffenegger$^{18}$,
R. Schwarz$^{23}$,
T.G. Tan$^{24}$,
A. H. M. J. Triaud$^{4}$,
Z. Benkhaldoun$^{25,26}$,
X. Bonfils$^{17}$,
\newauthor
F. Bouchy$^{8}$,
K. A. Collins$^{23}$,
F. Davoudi$^{6}$,
R. Doyon$^{9}$,
M. Gachaoui$^{26,6}$,
M. J. Hooton$^{27}$,
E. Jehin$^{28}$,
\newauthor
F. J. Pozuelos$^{29}$,
M. G. Scott$^{4}$,
S.~Yal\c{c}{\i}nkaya$^{6,30,31}$,
F. Zong Lang$^{18}$,
S. Z\'{u}\~{n}iga-Fern\'{a}ndez$^{6}$,
\newauthor
J. R. De Medeiros$^{32}$,
J. I. Gonz\'alez-Hern\'andez$^{5,33}$,
N. C. Santos$^{34,35}$
\\
\\
{\it A list of affiliations is given at the end of the paper.}
}

\date{Accepted 2026 February 26. Received 2026 February 24; in original form 2025 December 08}

\pubyear{\the\year{}}

\begin{document}
\label{firstpage}
\pagerange{\pageref{firstpage}--\pageref{lastpage}}
\maketitle

\begin{abstract}
We present the detection and validation of a small, temperate transiting exoplanet orbiting TOI-1080 every 3.9652482$_{-0.0000015}^{+0.0000014}$ days. The host is a quiet M4V star at 25.6~pc. The planet signal was first detected by \textit{TESS} and validated using \textit{TESS} and ground-based observations. By fitting the available light curves, the planet radius is measured to be 1.200$\pm$0.058R$_\oplus$ and its equilibrium temperature of $368_{-10.}^{+12}$~K. With NIRPS radial velocities, we are able to place a 3-$\sigma$ upper limit on the mass of TOI-1080~b of 10.7~M$_\oplus$. 
Our injection-recovery tests enable us to discard additional transiting planets in the TOI-1080 system with radii down to 0.9~R$_\oplus$ and periods between 0.5 and 7.7~days, and planets with radii larger than 1.4~R$_\oplus$ for periods up to 19~days. We demonstrate that it is highly amenable to characterisation of its mass and putative atmosphere. In particular, we find that TOI-1080~b is an exceptional target for the ongoing \textit{JWST+HST} Rocky Worlds DDT programme, having a priority score that is higher than four out of nine targets currently being investigated by the programme. 
TOI-1080~b can be added to the sample of nearby benchmark planets accessible for detailed study with \textit{JWST}. 
\end{abstract}




\section{Introduction}

Of the currently known over  6225
exoplanets \citep[][updated on  2026-02-05]{Schneider2011}, $\sim$75\% have been detected via the transit method, but of those, only $\sim$9\% are transiting stars cooler than 4\,000~K. 
But because M-dwarfs are the most numerous stars in the solar neighbourhood \citep{Henry2024, Kirkpatrick2024}, they could potentially harbour most of the terrestrial planets in the Galaxy \citep[e.g.,][]{Adams2005,Dressing2015, Burn2021, Mignon2025}. 
At first order, given that the transit depth is defined by the relative size between the stellar and planet radius \citep[e.g.,][]{Seager2003}, 
terrestrial planets \citep[R$_{\rm pl} \lesssim$ 1.6 R$_\oplus$; e.g.,][]{Rogers2015,Cloutier2020,Parviainen2024} are more easily detected and studied around these smaller stars. As such, the rocky planets most readily available for characterization now are those orbiting M-dwarf hosts. 
One of the stated science goals of the \textit{TESS} mission is to identify planets with radius $<$2.5 R$_\oplus$ around stars in the solar neighbourhood in orbits with periods of up to 10 days \citep{Ricker2014}. Currently, there are just over 183 of these nearby, small planets \citep[][updated on  2026-02-05]{exoplanetarchive}.

The growing sample of M-dwarf planets has shown that small planets orbiting M dwarfs can exhibit a wide-range of planet structures \citep[e.g.,][]{Dorn2017}. 
Although some rocky planets are expected to have atmospheres \citep[e.g.,][]{Zahnle2017, August2025}, a recent review by \citet{Kreidberg2025} highlights that there has been no unambiguous detection of an atmosphere of a rocky planet with the state-of-the-art \textit{JWST} observations. 
 The detections of a volatile-atmosphere around the hot Super-Earth 55~Cnc~e \citep{Hu2024},  a tenuous rock vapor atmosphere around K2-141~b  \citep{Zieba2022}, and a thick volatile-atmosphere around TOI-561~b \citep{Teske2025}, all orbit more massive stars. 
For example, the NIRISS transits of Super-Earth GJ~357~b show no atmospheric features \citep{Taylor2025} and the NIRSpec transits of  GJ~1156~b also show a flat transmission spectrum \citep{Bennett2025}. 
For TRAPPIST-1, thermal emission photometry with MIRI discarded dense compact atmospheres around the two inner planets \citep[][]{Gillon2025},
while transit spectroscopy with NIRISS and/or NIRSpec has failed to bring strong constraints on the presence/absence atmospheres of the five outer planets so far, because of limited signal-to-noise ratios  (SNR) and the negative impact of stellar noises \citep[e.g.,][]{Glidden2025}. 
The Rocky Worlds Discretionary Director's Time program has been allocated 500~h  with \textit{JWST} and 250 orbits with  \textit{HST} to expand our understanding of atmospheres of terrestrial planets orbiting M dwarfs
\citep{Redfield2024}. The first results from this program have found that for two secondary eclipses of GJ~3929~b observed with MIRI at 15$\mu$m, the most likely interpretation is a bare rock \citep{Xue2025}.

One of the significant hurdles to a robust atmospheric detection for these rocky M-dwarf planets is the stellar contamination due to their active planet hosts. When the stellar photosphere has inhomogeneities that are not occulted during transit, it leads to biases in the transmission spectra known as the transit light source effect \citep[e.g.,][]{Rackham2018}. Although strategies that rely on stellar models are currently applied, more development of stellar models are needed for robust mitigation \citep{Rackham2024}.
The atmospheric characterization of rocky planets can inform our understanding of their interior structure \cite[e.g.,][]{Lichtenberg2025}, internal sources of energy \citep[e.g.,][]{Lin2026} and give insights into the planet formation history \citep[e.g.,][]{Oberg2023}. As such, it has become important to detect and characterize small planets around small, inactive stars.

In this paper, we present the detection and validation of the planetary nature of TOI-1080~b, a temperate, rocky planet orbiting an inactive M4V star. 
We present first the characterization of the star TOI-1080 in Section~\ref{sec:star} based on medium resolution spectroscopy and spectral energy distribution fitting. We assess possible blends with archival images and high-resolution imaging, and discuss stellar activity. 
In Section~\ref{sec:alllcs}, we present the transit photometry obtained with \textit{TESS}  and the follow-up, time-series photometry obtained with ground-based facilities. 
 In Section~\ref{sec:validation}, we statistically evaluate the planet scenario against other astrophysical scenarios and validate the planet nature of TOI-1080~b.
In Section~\ref{sec:lcmodels}, we present the modeling of the transit light curves to characterize the physical properties of the planet TOI-1080~b.
The upper limit on the planet mass based on optical and near-infrared precision radial velocities is given in Section~\ref{sec:rvs}. 
We discuss the significance of this detection and the potential for further characterization of the planet in Section~\ref{sec:discussion}.


 \section{Stellar characterization of TOI-1080}\label{sec:star}

 As the physical properties of the planetary system depend on the stellar properties, here we present the careful stellar characterization of the host TOI-1080  based on optical (Section~\ref{sec:fire}) and near-infrared (Section~\ref{sec:magE}) medium-resolution spectra, and its spectral energy distribution (Section~\ref{sec:sed}). The star's coordinates, magnitudes and properties are summarized in Table~\ref{tab:starpar}.   We note that although TOI-1080 was observed with high-resolution  spectrographs (see Section~\ref{sec:rvs}), these observations do not have a sufficiently high SNR to derive accurate atmospheric parameters for the host star and are used only for radial velocities.

 \begin{table}
\centering
\begin{tabular}{@{}lp{25mm}p{30mm}@{}}
\toprule
{\bf Designations} & \multicolumn{2}{p{65mm}}{TOI-1080, TIC 161032923, 2MASS J18252886-5212514, APASS 25090057, Gaia DR2 6653968157663516672, Gaia DR3 6653968157663516672, UCAC4 189-186987, WISE J182528.54-521252.0, SIPS J1825-5212} \\ \midrule
{\bf Parameter} & {\bf Value}              & {\bf Source} \\ \midrule
T mag           & 12.707$\pm$0.007        & \cite{TICv8} \\
B mag           & 17.138$\pm$0.008      & \cite{ucac4} \\
V mag           & 15.349$\pm$0.049        & \cite{ucac4} \\
G mag           & 14.0343$\pm$0.0006      & \cite{gaiaDR3cat} \\
J mag           & 11.044$\pm$0.022          & \cite{2masscat} \\
H mag           & 10.467$\pm$0.023          & \cite{2masscat} \\
K mag           & 10.17$\pm$0.02         & \cite{2masscat} \\
W1 mag           & 10.015$\pm$0.023          & \cite{wisecat} \\
W2 mag           & 9.83$\pm$0.02          & \cite{wisecat} \\
W3 mag           & 9.584$\pm$0.042            & \cite{wisecat} \\
W4 mag           & $>8.42$       & \cite{wisecat} \\
Distance         & 25.559$\pm$0.054\,pc       & \cite{BJdist} \\
$\alpha$           & 18:25:28.4      & \cite{gaiaDR3cat} \\
$\delta$           & $-52$:12:52.46      & \cite{gaiaDR3cat} \\
$\mu_{\alpha}$           & $\rm -270.293\,mas\,yr^{-1}$      & \cite{gaiaDR3cat} \\
$\mu_{\delta}$           & $\rm -69.5518\,mas\,yr^{-1}$      & \cite{gaiaDR3cat} \\
Parallax      & 39.132 $\pm$ 0.022\,mas    & \cite{gaiaDR3cat} \\
SpT (Opt+NIR) & M4.0 $\pm$1.0                  & This work (\S\ref{sec:fire}-\ref{sec:magE})\\
$R_{\star}$     & $0.2019\pm 0.0075~\rm R_{\odot}$ & This work (\S\ref{sec:sed}) \\
$M_{\star}$     & $0.1667\pm0.0041~\rm M_{\odot}$   & This work (\S\ref{sec:sed})       \\
$L_{\star}$     & $0.00324\pm0.00046~\rm L_{\odot}$   & This work        \\
${\rm T_{eff}}$ & 3065$\pm$50 K             & This work                  \\
$\log g_\star$           & $5.049\pm0.043$           & This work                  \\
$\rm [Fe/H]$        & $-0.25\pm0.25$ dex              & This work (SED)   \\ 
$\rm [Fe/H]$        & $-0.09\pm0.11$ dex              & This work (near-infrared)   \\ 
$\rm [Fe/H]$        & $-0.10\pm0.20$ dex              & This work (optical)   \\ 
$\rm RV$    &     $-78.3\pm2.1$ $\rm km~s^{-1}$  &   \cite{gaiaDR3cat} \\
Age    &     $\gtrsim5-7$ Gyr  &  This work \\
         
\bottomrule
\end{tabular}
\caption{Stellar parameters adopted for this work.}
\label{tab:starpar}
\end{table}

 \subsection{Spectral type from near-infrared FIRE spectrum}\label{sec:fire}
We observed TOI-1080 with the FIRE spectrograph \citep{Simcoe2008} on the 6.5-m \textit{Magellan Baade} Telescope on UT 2021-10-15 under clear conditions with $0\farcs{7}$ seeing.
The observations were taken in the high-resolution echellette mode using the $0\farcs{6}$ slit, providing wavelength coverage of 0.82--2.51\,$\micron$ at a resolving power of $R{\approx}6000$.
We gathered four exposures of 200.8\,s at an average airmass of 1.2.
For flux and telluric calibration, we observed the A0\,V star HD\,163761 ($V{=}6.7$) in two 5-s exposures at a similar airmass.
ThAr arc lamp frames were collected at each pointing, along with daytime internal quartz and dome flats for pixel and blaze corrections.
The data were reduced with the {\sc FIREHOSE} pipeline\footnote{\url{https://github.com/rasimcoe/FIREHOSE}}.
The resulting spectrum has a median  SNR per resolution element of 360.

\begin{figure}
    \centering
    \includegraphics[width=\linewidth]{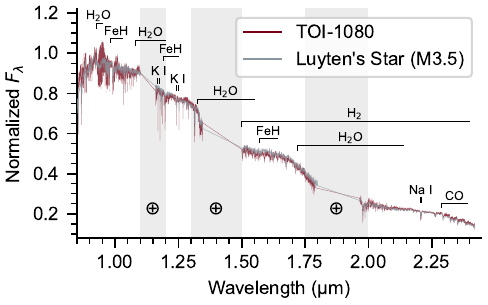}
    \caption{
        FIRE spectrum of TOI-1080.
        The target spectrum (red) is shown alongside that of the SpeX SXD spectrum of the M3.5\,V standard Luyten’s Star (GJ\,273; grey).
        The higher spectral resolution of the FIRE spectrum gives it a more jagged appearance.
        Strong M-dwarf spectral features and spectral regions with strong tellurics are indicated.
    }
    \label{fig:fire}
\end{figure}

The FIRE spectrum of TOI-1080 is shown in Fig.\,\ref{fig:fire}.
Comparison with single-star standards in the IRTF Spectral Library \citep{Cushing2005, Rayner2009} using the SpeX Prism Library Analysis Toolkit \citep[SPLAT;][]{splat} indicates the closest match to the M3.5\,V standard Luyten’s Star (GJ\,273), and we adopt a near-infrared spectral type of M3.5\,$\pm$\,0.5 accordingly.
From the $K$-band Na\,\textsc{i} and Ca\,\textsc{i} features and the H$_2$O--K2 index \citep{Rojas-Ayala2012}, we estimate $\mathrm{[Fe/H]} = -0.09 \pm 0.11$ using the \citet{Mann2013} relation. 
Furthermore, contrary to what is observed in giant stars, a visual inspection of the spectra of TOI-1080 reveals clear absorption by both the Na~I doublet at $\sim$2.21 $\mu$m and weak CO bands near 2.3 $\mu$m. This provides further support on the cool main-sequence nature of TOI-1080.

\subsection{Spectral type from optical MagE Spectrum}\label{sec:magE}

We observed TOI-1080 on UT 2023-05-22  with the Magellan Echellette spectrograph \citep[MagE;][]{Marshall2008} on the 6.5-m \textit{Magellan Baade} Telescope.
The night was clear, with seeing of $1\farcs3$.
We used the $0\farcs7 \times 10\arcsec$ slit, which provides a resolving power of $\lambda / \Delta \lambda \approx 6000$ over the 4000–9000\,\AA\ range.
Two 200-s exposures were obtained at an airmass of 1.1.
For flux calibration, we observed the spectrophotometric standard EG\,274 \citep{Hamuy1992, Hamuy1994} in two 10-s integrations.
Wavelength calibration and flat-field corrections were performed using ThAr arc frames at each pointing and the standard set of afternoon calibrations (biases, Xe flats, and internal flats).
Data reduction followed standard procedures using \textsc{PypeIt} \citep{pypeit_zenodo, pypeit_joss}.
The final spectrum has a  SNR of ${\sim}100$ per pixel at 8000\,\AA.

\begin{figure}
    \centering
    \includegraphics[width=\linewidth]{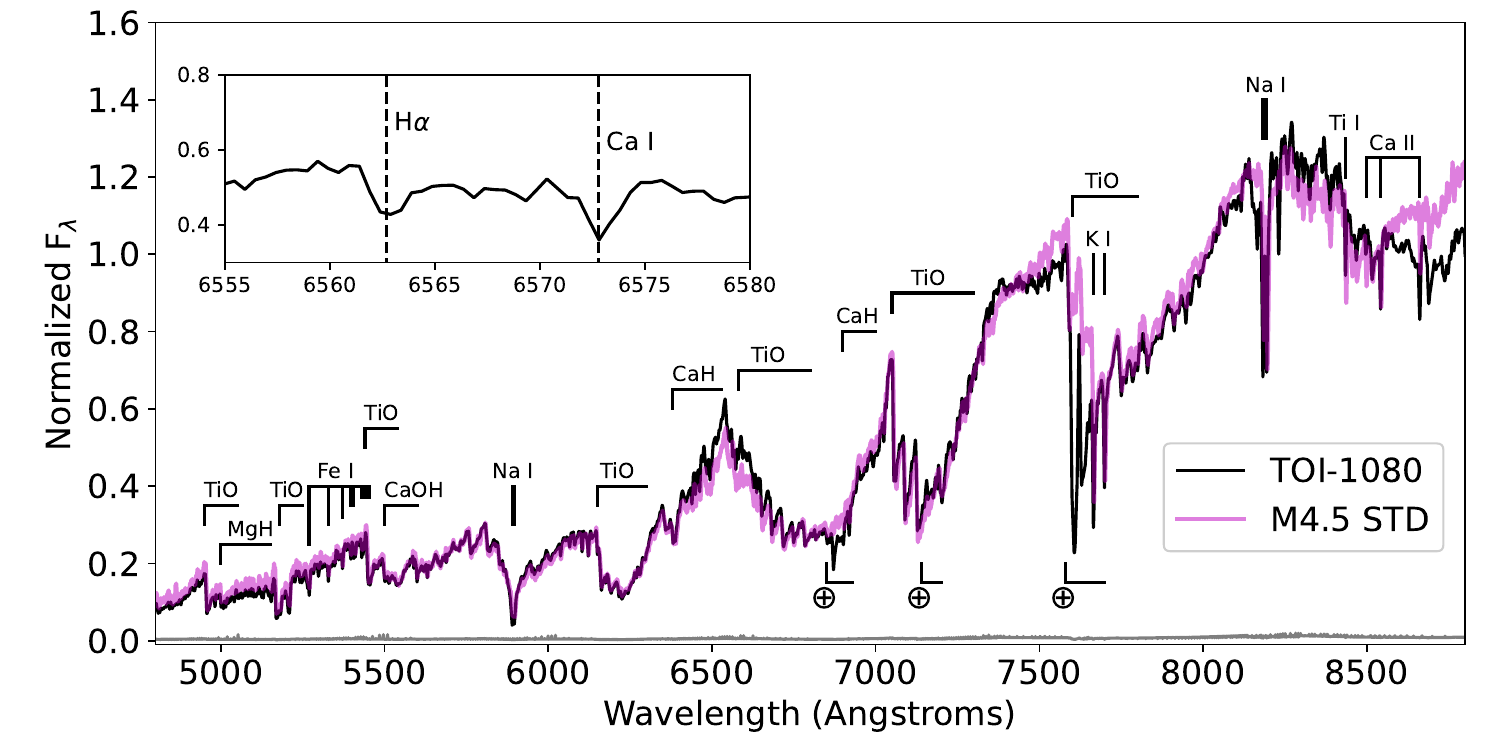}
    \caption{
        MagE optical spectrum of TOI-1080 (black line) normalized at 7500~{\AA} compared to a combined M4+M5 optical spectral template from \citet[][magneta line]{2007AJ....133..531B}. Standard M dwarf atomic and molecular features are labelled, as well as regions of uncorrected telluric absorption ($\oplus$). The inset box shows the 6555--6580~{\AA} region highlight absorption from H$\alpha$ and Ca~I.
    }
    \label{fig:mage}
\end{figure}

The reduced spectrum is shown in Figure~\ref{fig:mage}, and displays features typical of mid-type M dwarfs, with a red spectral slope; strong TiO and CaH molecular features; and line absorption from K~I, Na~I, Ca~I, Ca~II, Ti~I, and Fe~I. The spectrum appears to be intermediate between the M4 and M5 dwarf spectral templates from \citet{2007AJ....133..531B}, as confirmed by 
index-based classifications from
\citet{1995AJ....110.1838R,1997AJ....113..806G,2003AJ....125.1598L}; and \citet{2007MNRAS.381.1067R} which span M4--M4.5.
We adopt an optical classification of M4.5$\pm$0.5, slightly later but formally consistent with the near-infrared classification.
We detect the 6563~{\AA} H$\alpha$ feature in absorption with an equivalent width EW = 0.33$\pm$0.02~{\AA}, indicating weak or absent magnetic emission and hence an activity age $\gtrsim$5--7~Gyr \citep{2008AJ....135..785W}.
The metallicity index $\zeta$ = 0.909$\pm$0.002 \citep{2007ApJ...669.1235L,2013AJ....145..102L} indicates a near-solar metallicity of [Fe/H] = $-$0.1$\pm$0.2 based on the calibration of \citet{Mann2013}, similar to that inferred from the near-infrared spectrum.

\subsection{Spectral Energy Distribution for TOI-1080}\label{sec:sed}

\begin{figure}
    \includegraphics[width=\columnwidth,trim=2.3cm 2.6cm 2.1cm 3.1cm,clip]{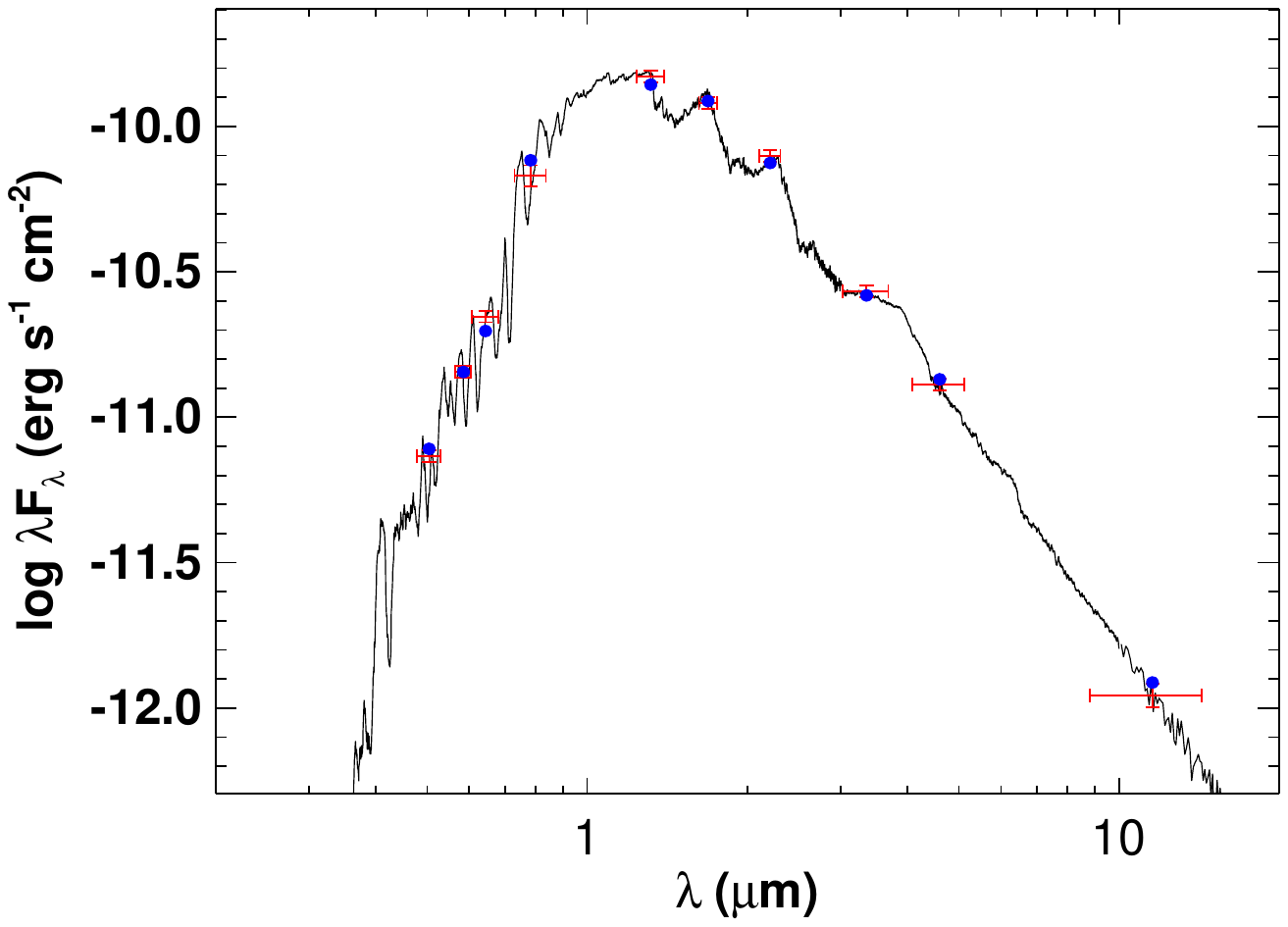}
    \caption{Spectral energy distribution of TOI-1080. Red symbols represent the observed photometric measurements, where the horizontal bars represent the effective width of the passband. Blue symbols are the model fluxes from the best-fit PHOENIX atmosphere model (black).  \label{fig:sed}}
\end{figure}

 As an independent determination of the basic stellar parameters, we performed an analysis of the broadband spectral energy distribution (SED) of the star together with the {\it Gaia\/} DR3 parallax \citep[with no systematic offset applied; see, e.g.,][]{StassunTorres:2021}, in order to determine an empirical measurement of the stellar radius, following the procedures described in \citet{Stassun:2016,Stassun:2017,Stassun:2018}. We pulled the $JHK_S$ magnitudes from {\it 2MASS}, the $G_{\rm BP} G_{\rm RP}$ magnitudes from {\it Gaia}, the $gri$ magnitudes from {\it APASS}, and the W1--W3 magnitudes from {\it WISE}. Together, the available photometry spans the full stellar SED over the wavelength range 0.4--10~$\mu$m (see Fig.~\ref{fig:sed}).  
 
We performed a fit using PHOENIX stellar atmosphere models \citep{Husser:2013}, with the free parameters being the effective temperature ($T_{\rm eff}$) and metallicity ([Fe/H]). The extinction, $A_V$, was fixed at zero due to the proximity of the system. The resulting fit (Fig.~\ref{fig:sed}) has a reduced $\chi^2$ of 1.6, with a best-fit $T_{\rm eff} = 3065 \pm 50$~K, and [Fe/H] $= -0.25 \pm 0.25$. Integrating the model SED gives the bolometric flux at Earth, $F_{\rm bol} = 1.589 \pm 0.056 \times 10^{-10}$ erg~s$^{-1}$~cm$^{-2}$. Taking the $F_{\rm bol}$ together with the {\it Gaia\/} parallax directly yields the stellar radius via the Stefan-Boltzmann relation, giving $R_\star = 0.2019 \pm 0.0075$~R$_\odot$. In addition, we can estimate the stellar mass from the empirical relations of \citet{Mann2019}, giving $M_\star = 0.1667\pm0.0041$~M$_\odot$.

\subsection{Search for blends}\label{sec:blends}

\subsubsection{SOAR speckle imaging}\label{sec:highresim}
\begin{figure}
    \includegraphics[width=\columnwidth]{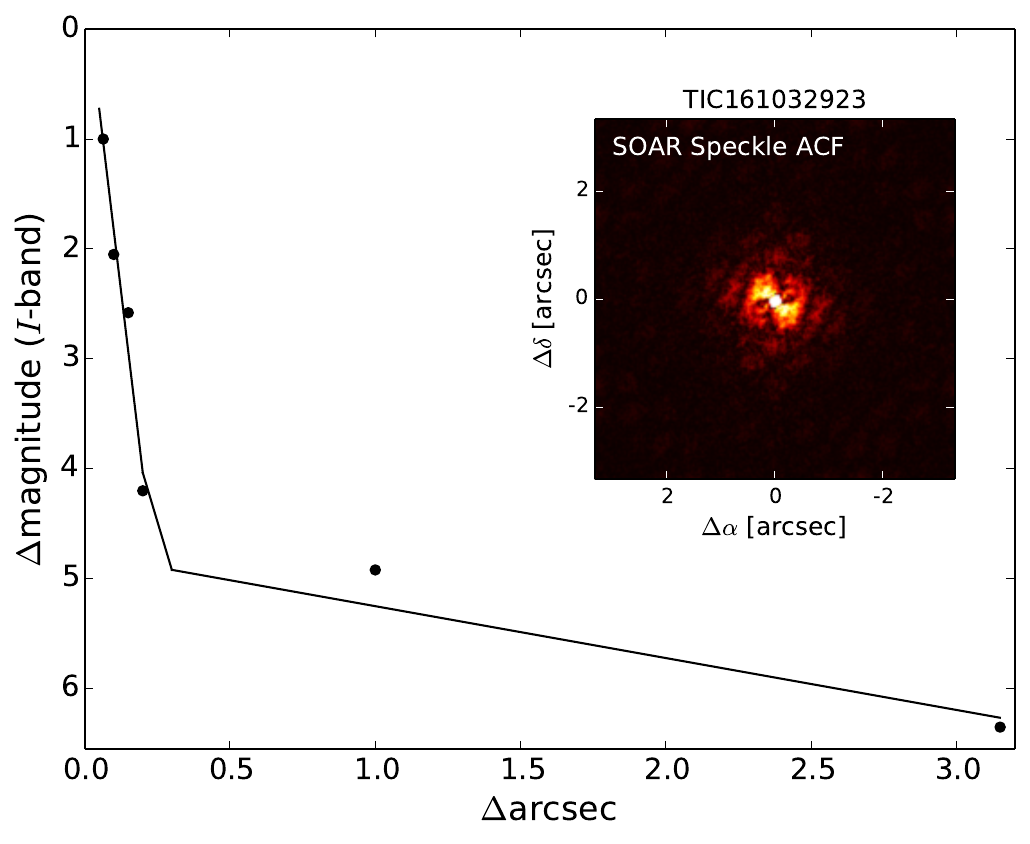}
    \caption{Detection sensitivity and auto-correlation (inset) functions for the TOI-1080 SOAR speckle imaging.
    We do not identify any nearby stars within 3$\arcsec$ of TOI-1080. 
    }
    \label{fig:highres}
\end{figure}

High-angular resolution imaging is needed to search for nearby sources that can contaminate the \textit{TESS} photometry, resulting in an underestimated planetary radius, or be the source of astrophysical false positives, such as background eclipsing binaries. We searched for stellar companions to TOI-1080 with speckle imaging on the 4.1-m Southern Astrophysical Research (SOAR) telescope \citep{Tokovinin2018} 
on UT 2021-07-14, observing in Cousins I-band, a similar visible bandpass as TESS. This observation was sensitive with 5-$\sigma$ detection to a 5.0-magnitude fainter star at an angular distance of 1$\arcsec$ from the target. More details of the observations within the SOAR \textit{TESS} survey are available in \citet{Ziegler2020}. 
The 5-$\sigma$ detection sensitivity and speckle auto-correlation functions from the observations are shown in Fig.~\ref{fig:highres}. No nearby stars were firmly detected within 3$\arcsec$ of TOI-1080 in the SOAR observations.

\subsubsection{Archival images of TOI-1080}\label{sec:archival}

\begin{figure*}
\includegraphics[width=1.0\textwidth,trim=0.27cm 0.27cm 0.15cm 0.27cm, clip]{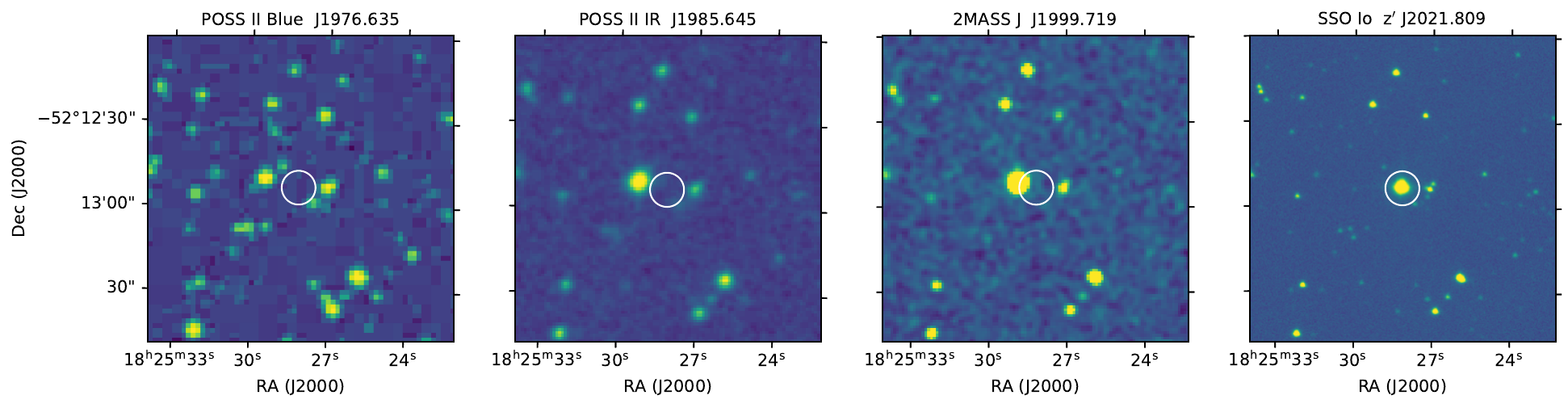}
      \caption{Archival images for TOI-1080 from the Digital Sky Survey (DSS) and 2MASS compared to the SSO Io observations taken in 2021. 
        The position of TOI-1080 in the 2021 SSO Io stack image is marked in all panels by the circle and the coordinates are given in J2000.  
      All images are the same size of 0.03$^{\circ} \times$0.03$^{\circ}$, and are centered on TOI-1080's position in the 2021 SSO Io observations. The color scale for each frame has been optimized to better show the stars. 
      }
         \label{fig:archival}
   \end{figure*}

Figure~\ref{fig:archival} shows the archival images for the TOI-1080 field. We have  centered TOI-1080 in the SPECULOOS-Southern Observatory (SSO) Io image taken in 2021 and note that the position is not significantly different in the SSO Europa image taken in 2025. Over the span of over 45 years that the archival images encompass, there is no other star in the position of the SSO Io observations of TOI-1080. Thus, it is unlikely that there is contamination by background source that is bright enough to affect our photometry and results.

\subsection{Stellar Activity}\label{sec:activity}
We analyzed the \textit{TESS} and SPECULOOS light curves, described in the next section, of TOI-1080 to search for evidence of stellar rotational modulation or flaring. We could not identify any flares in the light curves or any  rotationally-induced photometric variability. A frequency analysis of the \textit{TESS} photometry  (i.e., both PDCSAP and SAP fluxes), with in-transit data masked, was performed using a generalized Lomb--Scargle periodogram \citep[GLS;][]{Zechmeister2009} to search for the stellar rotation period. When analyzed independently, each of the three sectors yields a different period  and the phased light curves do not show clear rotational modulation, 
suggesting that the signals are not significant.
 The TESS SAP light curves present some variability that if it is due to rotation, it has a much longer timescale than any given sector. Because the sectors are too far apart, the modulation is not coherent on any particular period. We could not derive a robust rotation period from the TESS data set, indicating that any spot modulation of the light curves has a small amplitude and/or a long rotation period below our detection sensitivity.
 We further analysed the ground-based photometry from SPECULOOS, which comprises 243~hours of observations spanning $\sim$1364 days. The resulting periodogram is noisy, with the amplitude at the period of maximum power remaining below 0.4~ppt. The low amplitude of photometric variability and absence of flares in both datasets indicate that TOI-1080 is a quiet star.

 Unfortunately, the available high-resolution spectra of TOI-1080 do not have sufficient signal in the \ion{Ca}{II}\,H\&K lines (SNR $<$1) nor in H-$\alpha$ (SNR $\sim$5) to assess chromospheric activity via spectroscopic indicators.

\section{Transit photometry: observations and data reduction}\label{sec:alllcs}

In this section, we describe the observations and data reduction used in the identification of the transit events in the \textit{TESS} light curves and the characterization of the transiting planet TOI-1080~b with additional ground-based observations. 

\subsection{TESS light curve and planet candidate status} \label{sec:tesslc}

\begin{figure*}
    \centering
    \includegraphics[width=0.8\textwidth]{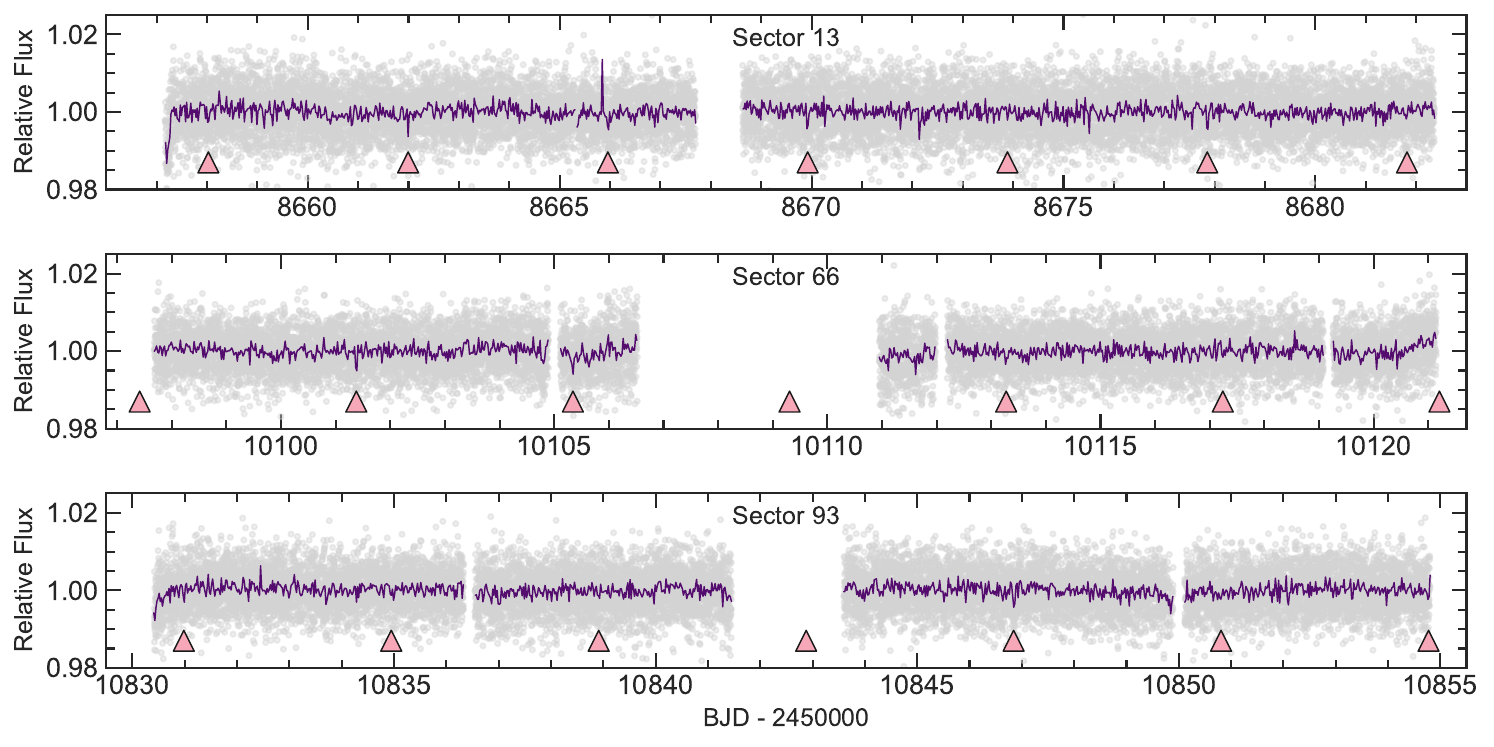}
    \caption{Normalized \textit{TESS} light curve. The 45\,967 individual photometric points are shown in grey. The purple line shows the 30~minute binned flux. The transits of TOI-1080~b are marked by the pink triangles.}
    \label{fig:tess}
\end{figure*}


TOI-1080 was observed by \textit{TESS} in Sector 13 (2019-04-22 to 2019-05-21), 66 (2023-06-02 to 2023-07-01), and 93 (2025-06-03 to 2025-06-29) with a 2-minute cadence. 
The Science Processing Operations Center \citep[SPOC;][]{Jenkins2016} NASA Ames Research Center conducted a transit search in Sector 13 using a wavelet-based, noise-compensating matched filter \citep{TPS2002ApJ...575..493J,TPS2010SPIE, 2020TPSkdph}
and detected a threshold crossing event. 
The signature passed all the tests in the Data Validation (DV) report \citep{Twicken:DVdiagnostics2018, Li:DVmodelFit2019}, 
including the difference image centroid test, which constrained the location of the transit source to within 4.3 $\pm$ 5.0 arcsec of the catalog location of TOI-1080. The \textit{TESS} Science Office reviewed the data validation reports and issued an alert for TOI-1080 b on UT 2019 August 16 \citep{guerrero:TOIs2021ApJS}. The transit signature was detected in a subsequent search of sectors 13 and 66 where the difference image centroid test constrained the transit source to within 5.3 $\pm$ 3.0 arcsec. 
The three-sector \textit{TESS} light curve is composed of 45\,967 individual photometric points and it spans $\sim$2\,198 days. 

SPOC data products are publicly available as simple aperture photometry \citep[SAP,][]{twicken:PA2010SPIE,morris:PA2020KDPH} or presearch data conditioning simple aperture photometry \citep[PDCSAP,][]{Stumpe2012,Smith2012,Stumpe2014}; the latter is corrected for instrumental systematics and contamination of known nearby sources. We downloaded the PDCSAP lightcurve for TOI-1080 from the NASA Milkulski Archive for Space Telescopes via {\sc LightKurve} \citep{lightkurve}. The \textit{TESS} lightcurve for all three sectors is presented in Fig.~\ref{fig:tess}.

\begin{figure*}
    \centering
    \includegraphics[width=0.315\textwidth]{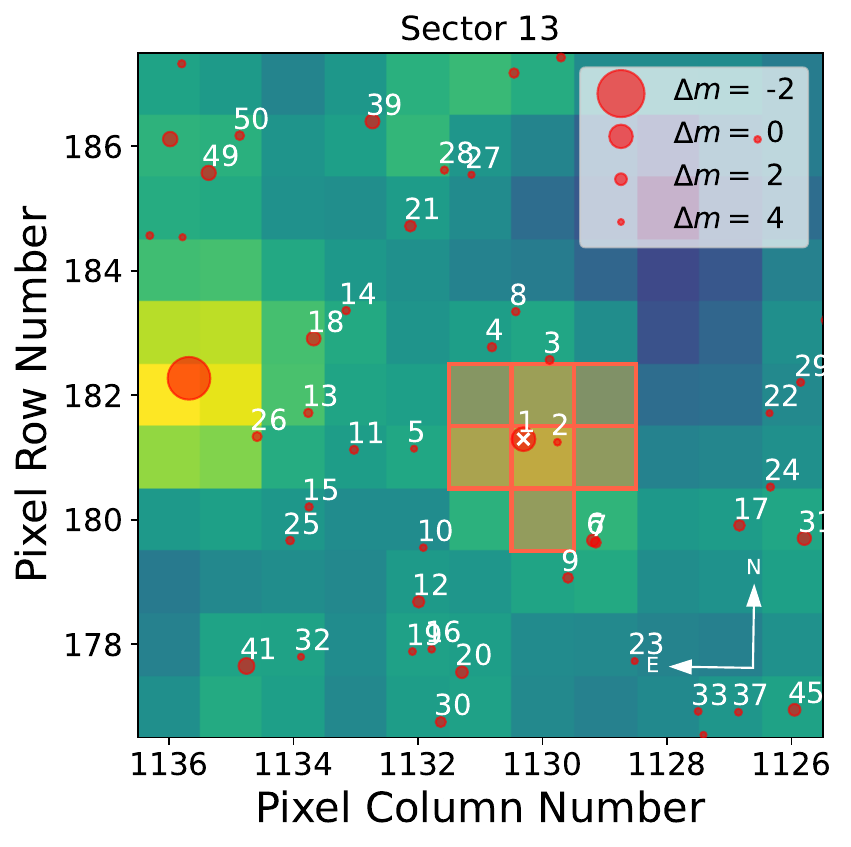}
    \includegraphics[width=0.315\textwidth]{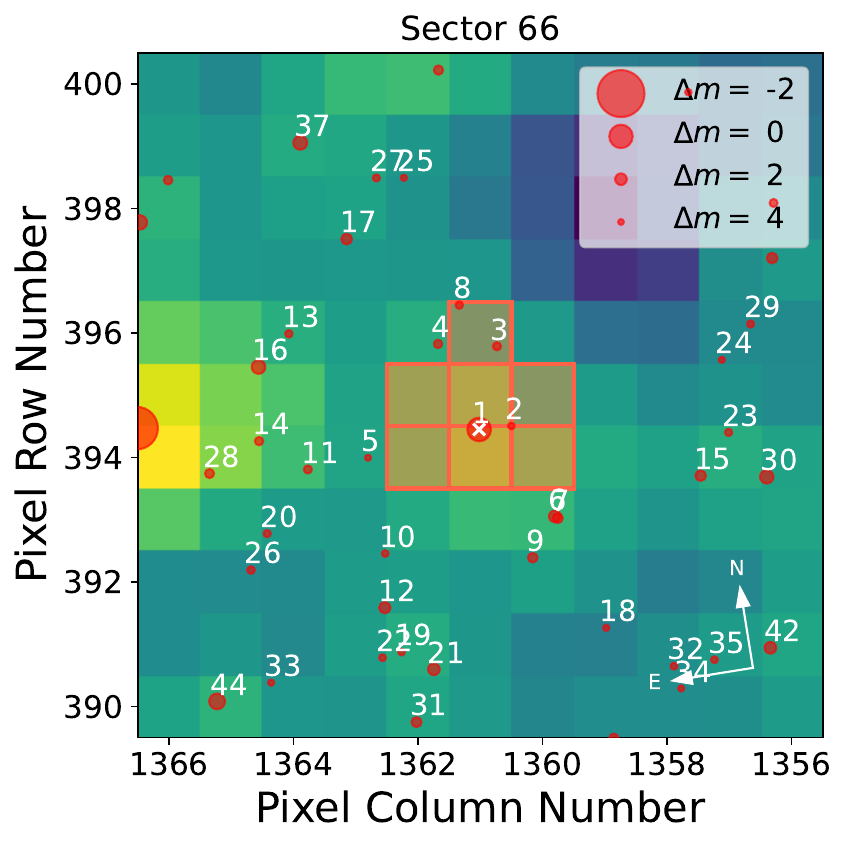}  \includegraphics[width=0.346\textwidth,height=5.6cm]{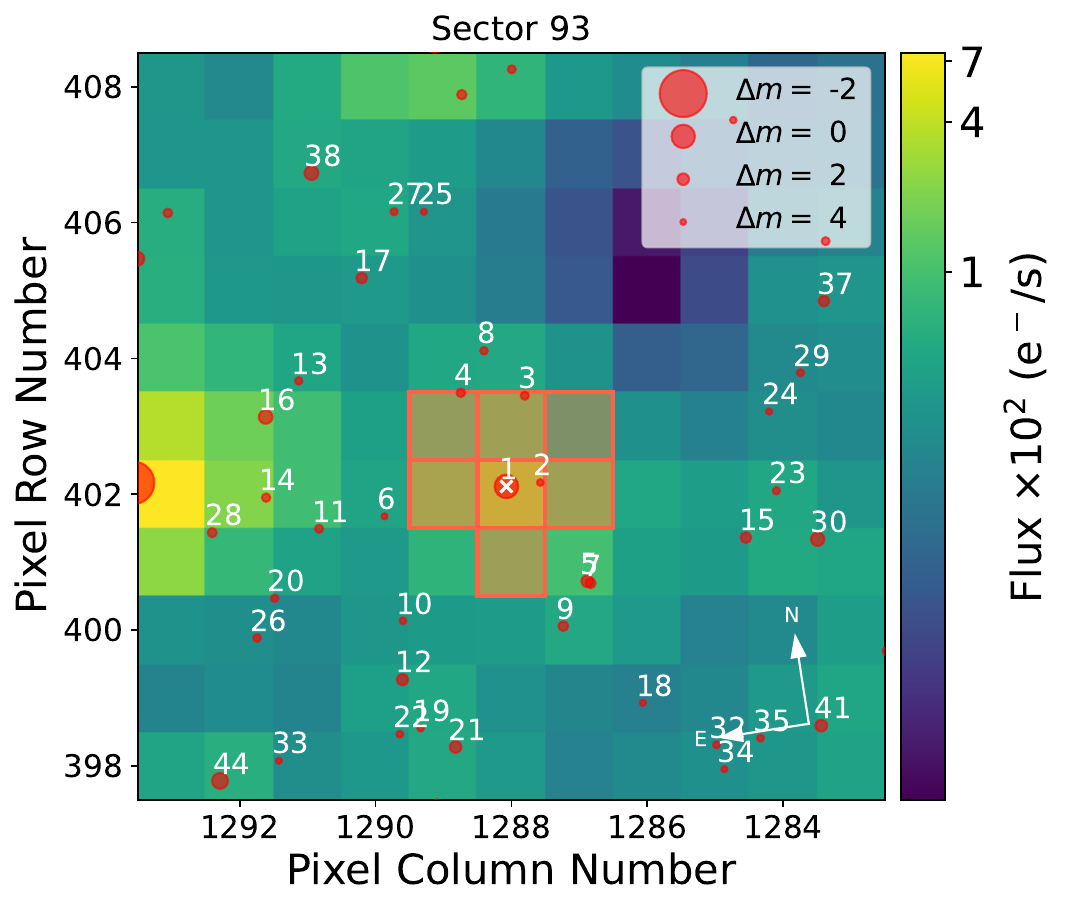}
    \caption{TESS Target pixel files and apertures (shaded in red). Although, the TOI-1080 field is relatively crowded, the brightest star is not affecting the \textit{TESS} photometry and nearest stars are $>$4 magnitudes fainter than the target in the \textit{TESS} passband. 
    }
    \label{fig:tpf}
\end{figure*}

With the python package \textsc{tpfplotter} \citep{Aller2020}, we plotted the field of view and aperture of the three \textit{TESS} sectors to assess contamination from nearby stars using the Gaia~DR3 catalog \citep{gaiaDR3cat,gaiadr3} and to discard that the source of the transit event itself in the \textit{TESS} light curve is another star in the field. Figure~\ref{fig:tpf} shows the target pixel files for each \textit{TESS} sector from which we can conclude the dilution is minimal in the \textit{TESS} light curves from nearby stars.

\begin{table*}               
\centering          
\begin{tabular}{l c c c c l }     
\hline\hline       
Date (UT) & Facility & Filter & $t_{\rm exp} (s) $ & $\Delta\ln{Z}$& Notes\\
\hline
2019-09-02 & LCO-SAAO-1.0m & $Sloan-i'$ & 100 &   8.32     & full \\ 
2020-05-09 &  LCO-SSO-1.0m & Pan-STARRS $z_s$ & 150 &   0.79    &full \\
2020-05-21 & PEST & $R_{\rm C}$ & 60 &   -3.04      & partial \\
2020-08-01 &  LCO-SAAO-1.0m  & $Sloan-r'$ & 150 &    6.48     &full \\ 
2020-08-16 & LCO-SAAO-1.0m & Pan-STARRS $z_s$ & 150 &   11.89    &full \\
2020-08-16 & LCO-SAAO-1.0m & $V$ & 150 &   0.72   &full \\
2020-08-24 & LCO-SAAO-1.0m & Pan-STARRS $z_s$ & 150 &   3.04    &full \\
2021-04-19 &LCO-SSO-1.0m  & $Sloan-i'$ & 100  &   229.59    & full \\
2021-05-24 & ExTrA 2 \& 3$^\star$ &  1.21 $\mu$m & 60  &   --     & full \\
2021-06-01 & ExTrA 2 \& 3 & 1.21 $\mu$m & 60  &    --    & full \\
2021-06-05 & TRAPPIST-South & $I+z$ & 65  &   20.78    & full \\
2021-06-05 & ExTrA 2 \& 3 & 1.21 $\mu$m & 60  &    --    & full \\
2021-06-09 & ExTrA 2 \& 3 & 1.21 $\mu$m & 60  &    --    & full \\
2021-10-18 & ExTrA 2 \& 3 & 1.21 $\mu$m & 60   &   --    & full \\
2021-10-22 & SPECULOOS Io & $Sloan-z'$ & 10  &  0.42  & full\\
2021-10-22 & ExTrA 2 \& 3 & 1.21 $\mu$m & 60  &   --     & full \\
2021-10-26 & TRAPPIST-South & $I+z$ & 65  &   -1.88    & partial \\
2022-05-08 & ExTrA 2 \& 3 & 1.21 $\mu$m & 60   &   --    & full \\
2022-05-12 & ExTrA 2 \& 3 & 1.21 $\mu$m & 60   &   --    & full \\
2023-04-14 & SPECULOOS Io & $Sloan-z'$ & 10  &   25.81    & full\\ 
2023-08-23 & ExTrA 2 \& 3 & 1.21 $\mu$m & 60   &   --    & full \\
2023-08-27 & ExTrA 2 & 1.21 $\mu$m & 60  &    --    & full \\
2024-08-14 & ExTrA 2 & 1.21 $\mu$m & 60   &   --    & full \\
2025-06-27 & SPECULOOS Europa & Sloan-$z'$ & 10  &   31.046    & full \\
\hline  
\end{tabular}
\caption{Ground-based time-series photometric follow-up observations of TOI-1080. The $\Delta \ln Z$ column is the Bayesian evidence for the transit, as described in Section \ref{sec:lcmodels}. Only lightcurves with $\Delta \ln Z > 5$ are used in the global analysis. $\star$: The ExTrA lightcurves were treated as a single long time series for each telescope as described in Section \ref{sec:lcmodels}; we therefore do not report a Bayes Factor for each individual lightcurve.}
\label{tab:lcs}
\end{table*}

\subsection{Follow-up, ground-based light curves} \label{sec:lcs}

To better constrain the planet radius and ephemeris, we obtained ground-based, time-series photometry for the transit of TOI-1080~b.  A summary of the follow-up, time-series observations is presented in Table~\ref{tab:lcs}. 
 
\subsubsection{SPECULOOS photometry}\label{sec:lcspc}
We made use of two of the four telescopes of the SPECULOOS-Southern Observatory \citep[SSO;][]{2018_gillon_spc,SPC_Laeti,Sebastian2021} to obtain transit light curves of TOI-1080~b. SSO is located at ESO Paranal Observatory, and is made up of four identical 1m-class telescopes named after the Galilean moons (Io, Europa, Ganymede, and Callisto). They are equipped with a $2\mathrm{K}\times2\mathrm{K}$ Andor iKon-L cameras covering a field of view of $12\arcmin \times12\arcmin$ for a pixel scale of 0\farcs35. We obtained three full transits of TOI-1080.01: two with the Io telescope on UT 2021-10-22 and UT 2023-04-14, and one the Europa telescope on UT 2025-06-28 in the \textit{Sloan-z'} filter and with 10s exposures. We reduced the data and performed differential photometry using a custom pipeline built with the \textsc{prose} package \citep{prose_2022}. 

In addition to transit observations, we obtained 27 nights of continuous monitoring of the star between UT 2023-06-26 and UT 2023-07-26. The goal was to search for hints of other possible candidates in the system, and assess our performance and the detectability of such candidates with the SPECULOOS telescopes. These observations were carried out in the \textit{Sloan-z'} filter with exposures of 10s, and they make a total of 243~hours. 

\subsubsection{LCO-SAAO and LCO-SSO photometry} \label{sec:lcslco}

We used the Las Cumbres Observatory Global Telescope (LCOGT; \citealt{Brown_2013}) 1.0m network to observe a total of eight full transits of TOI-1080\,b.
Seven transits were observed with the LCOGT 1.0 telescope at the South Africa Astronomical Observatory (LCO-SAAO) and one was observed with the LCOGT 1.0 telescope at Siding Spring Observatory (LCO-SSO) in Australia.
The telescopes are equipped with a $4096\times4096$ SINISTRO camera having an image scale of $0\farcs389$ per pixel and a field of view of $26\arcmin\times26\arcmin$.

The observations were carried out in the Sloan-$r'$, Sloan-$i'$, Johnson-$V$ and Pan-STARRS $z_s$ filters with exposure times of 150s, 100s, 150s, and 150s, respectively. The observation dates are given in Table\,\ref{tab:lcs}.
The data calibration was performed using the standard LCOGT {\sc BANZAI} pipeline \citep{McCully_2018SPIE10707E}.
The photometry extraction was performed in an uncontaminated 3.1--5.2\arcsec target apertures using {\sc AstroImageJ} \citep{Collins_2017}.

\subsubsection{TRAPPIST-South photometry}\label{sec:lcstrap}

TRAPPIST-South  \citep[TRAnsiting Planets and PlanetesImals Small Telescope,][]{Jehin2011,Gillon2011} is a 60cm Ritchey-Chr\'etien robotic telescope located at La Silla Observatory in Chile since 2010. The telescope is equipped with a 2K$\times$2K FLI Proline CCD camera with a FOV of $22\arcmin\times22\arcmin$ and a pixel scale of 0.65\arcsec/pixel.
Two full transits were observed on UT 2021-06-05 and UT 2021-10-26 in the $I+z$ filter with exposure time of 65s. The data reduction and photometric extraction were performed using the \textit{PROSE} pipeline \citep{prose_2022}.

\subsubsection{ExTrA photometry}

ExTrA is an instrument composed of three 60-cm telescopes located at La Silla Observatory in Chile that is capable of obtaining near-infrared photometry \citep{Bonfils2015}. 
We have observed TOI-1080 during 12 nights using one or two of the ExTrA telescopes between UT 2021-05-24 and UT 2024-08-14. We used 8\arcsec aperture fibers and the lowest-resolution mode (R$\sim$20) of the spectrograph with an exposure time of 60~seconds.
The details of the observations are found in Table~\ref{tab:lcs}.  
To process the data, we followed the procedure described in detail in \citet{Cointepas2021} to subtract the sky flux and normalize by the flat, and 
obtained the differential photometry for TOI-1080 using the comparison stars. 

\subsubsection{PEST photometry}\label{sec:lcpest}

The Perth Exoplanet Survey Telescope (PEST) has a 0.3-m diameter telescope and is located near Perth, Australia. PEST observed the partial transit of TOI-1080~b on UT 2020-05-21 with 60s exposures in the R$_{\rm C}$. A total of 315 images over 401 mins were obtained with a $1530\times1020$ SBIG ST-8XME camera with an image scale of 1$\farcs$2 pixel$^{-1}$ resulting in a $31\arcmin\times21\arcmin$ field of view. A custom pipeline based on {\sc
C-Munipack}\footnote{http://c-munipack.sourceforge.net} was used to
calibrate the images and extract the differential photometry.

\section{Planet validation} \label{sec:validation}



We made use of the statistical validation software \textsc{TRICERATOPS} \citep{triceratops_soft,triceratops} to test the probability of the transiting planet hypothesis for the TOI-1080.01 transit signal. \textsc{TRICERATOPS} calculates the flux contribution from nearby stars to assess whether they could be responsible for the transit signal. It then calculates probabilities for a variety of transiting planet and eclipsing binary scenarios on the target and nearby stars by fitting lightcurve models, and finally provides a false positive probability (FPP). The typical threshold for statistical validation of a planet is an FPP of 0.015.

In the first instance, we calculated the FPP using the phase-folded \textit{TESS} photometry, including all three sectors, and the contrast curve from our high-resolution imaging described in Section \ref{sec:highresim}; the resulting value for the FPP is 0.01626 which is just above the threshold. 

We then used {\sc TRICERATOPS}' functionality that allows a user to discard certain false positive scenarios if ground-based observations have already ruled them out. We dropped NTP (nearby transiting planet), NEB (nearby eclipsing binary) and NEBx2P (nearby eclipsing binary at twice the period) for 13 nearby stars, as the small pixel scale of our ground-based photometry conclusively rules out these stars are the source of the transit event. This reduced the FPP to 0.001948, placing it confidently below the threshold for statistical validation.

We therefore consider the planet to be validated and refer to it as TOI-1080~b.

\section{Light curve modeling} \label{sec:lcmodels}

\begin{figure}
    \centering
    \includegraphics[width=\columnwidth]{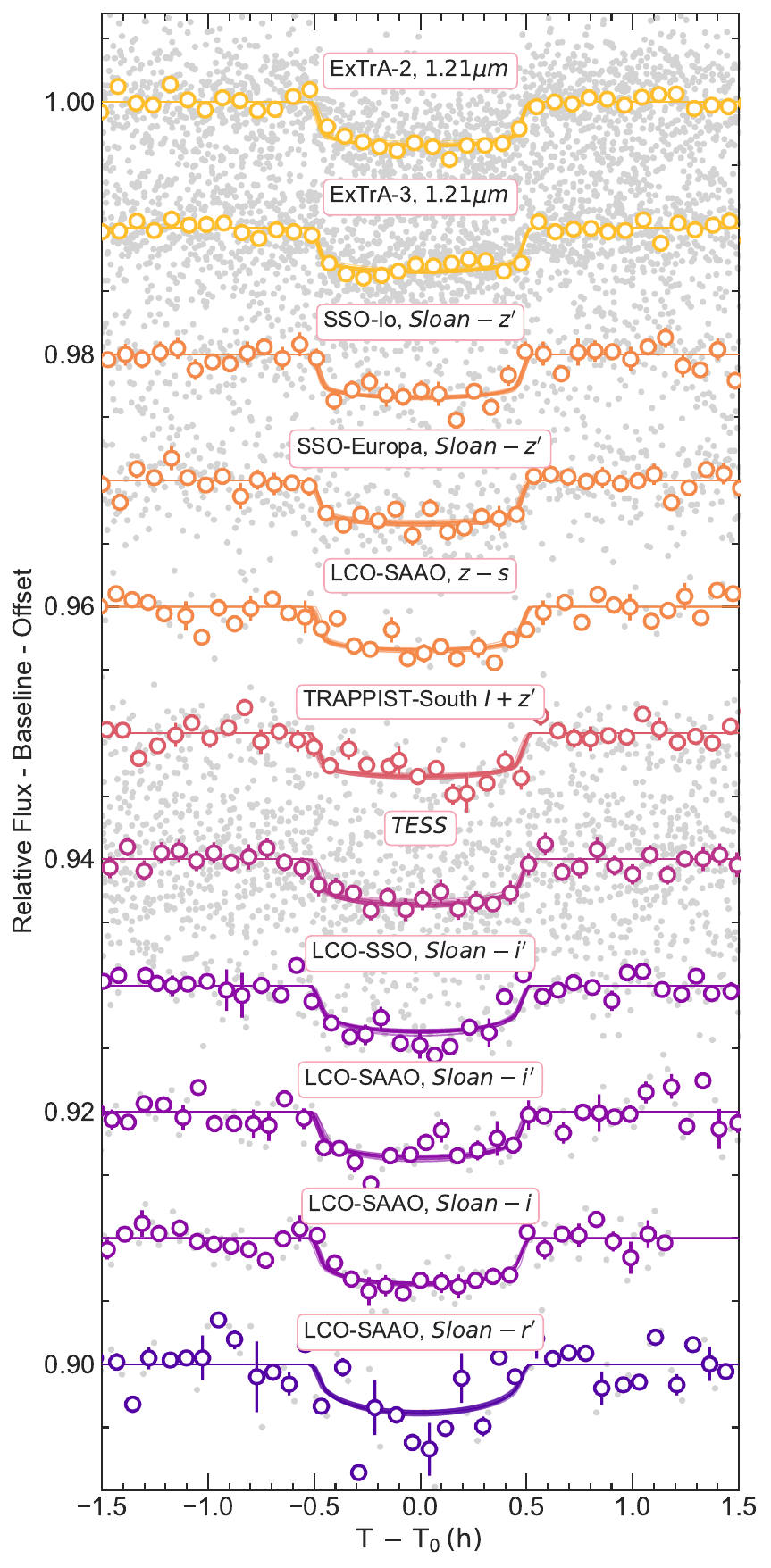}
    \caption{Transit photometry of TOI-1080~b along with best fitting models. Grey points are the raw flux while white circles are binned to 15~minutes, coloured according to the filter used in the observation. Photometry from \textit{TESS} and ExTrA are phase folded. The model lines are 20 random draws from the posterior. }
    \label{fig:transits}
\end{figure}
 
We carried out a global analysis of the complete photometric dataset described in Section \ref{sec:alllcs} using \textsc{Allesfitter} \citep{AllesfitterSoft,AllesfitterPaper}, a flexible \textsc{Python} inference package. \textsc{Allesfitter} generates lightcurve models using \textsc{Ellc} \citep{ellc}, and best-fitting models can be chosen using either nested sampling or MCMC algorithms, implemented via \textsc{Dynesty} \citep{dynesty} and \textsc{Emcee} \citep{emcee} respectively. For our analysis we used the nested sampling algorithm as this allows effective model comparison by calculating the Bayes Factor \citep{BayesFactor}, enabling us to choose the model with the highest statistical evidence. 

We adopted the signal parameters reported by SPOC\footnote{Available at \url{https://exofop.ipac.caltech.edu/tess/target.php?id=161032923}} as uniform priors, and the stellar parameters adopted in Section \ref{sec:star} as normal priors. Additionally, we used \textsc{PyLDTK} \citep{pyldtk} and the Phoenix stellar atmosphere models \citep{phoenix} to calculate quadratic limb-darkening coefficient priors for each photometric band used in our follow-up campaign. We reparametrised these in the parametrisation of \cite{kippingldcs} and adopted them as normal priors, following e.g., \cite{Dransfield2022,Triaud2023,Dransfield2024,Scott2025}. 

Several of the transits described in Section \ref{sec:lcs} are dominated by red noise. To assess which lightcurves could contribute to a refinement of the transit parameters, we follow the procedure thoroughly described in \cite{pozuelos2023} and subsequently applied in \cite{2267}. In brief, we fit two models to each individual ground-based lightcurve; the first is a pure noise model with fixed transit parameters, including $R_{\rm p}/R_\star$=0, and the second is a noise+transit model. For the noise model we use \textsc{Allesfitter}'s hybrid spline functionality. We then compare the Bayesian evidence between the pure noise and the transit+noise models, and select only those with $\Delta \ln Z>5$ to include in the final model from which the planet's physical properties are derived.  The derived value of $\Delta \ln Z$ for each lightcurve is presented in the penultimate column of Table \ref{tab:lcs}.

In the case of the lightcurves obtained with ExTrA, we stitched together the lightcurves from each of the two telescopes used. In both cases, the transit+noise model was significantly preferred; however, the photometry has very large systematics and are difficult to detrend using the hybrid spline. We therefore use a Gaussian Process (GP) kernel to model the red noise only in the case of the ExTrA lightcurves. We select the simple harmonic oscillator (SHO) kernel as implemented by {\sc Celerite} \citep{celerite}. When detrending with a GP, the main challenge is to avoid overfitting; to avoid this we place a uniform prior on the GP frequency ($\omega_{0}$) such that only low frequency noise is removed. We also remove three individual transits which have high amplitude trends that cannot be removed with our frequency upper limit. 

We then jointly fit the \textit{TESS} photometry and the highest quality ground-based lightcurves, fitting in the first instance a circular model and one with free eccentricity. In both cases we fit for all transit parameters ($R_{\rm p}/R_\star$, $(R_\star+R_{\rm p})/a$, $\cos\,i$, $T_0$, and $P$). In the free eccentricity model, we parametrise the eccentricity and argument of periastron as $\sqrt{e_{\rm b}} \cos{\omega_{\rm b}}$ and $\sqrt{e_{\rm b}} \sin{\omega_{\rm b}}$.

The eccentricity derived from the more complex model is $e_\mathrm{b}=0.133_{-0.095}^{+0.23}$; this value is consistent with zero at the $2\sigma$ level.
This small eccentricity has a 37.5\% probability of being spurious, as defined by \citet{Lucy1971}. 
Additionally, we compare the statistical evidence for the circular and free eccentricity models by calculating the log Bayes factor, and find that the circular model is very marginally preferred ($\Delta \ln Z\approx0.3$). We therefore adopt the circular model and we report all fitted and derived parameters in Table \ref{tab:global_fit}. In Fig.~\ref{fig:transits} we present all fitted transit lightcurves along with best fitting models. 

\begin{table}
    \centering
    {\renewcommand{\arraystretch}{1.2}
    \begin{tabular}{ccc}
\hline 
\multicolumn{3}{c}{\textit{Priors and Fitted parameters}} \\ 
\hline 
$R_b / R_\star$ & $\mathcal{U}(0.04,0.07)$& $0.0548\pm0.0012$  \\ 
$(R_\star + R_b) / a_b$ & $\mathcal{U}(0.02,0.07)$ & $0.0360_{-0.0016}^{+0.0020}$ \\ 
$\cos{i_b}$ & $\mathcal{U}(0.0,0.03)$ & $0.0129_{-0.0058}^{+0.0047}$ \\ 
$T_{0;b}$$^{\dagger}$, ($\mathrm{BJD}$) & $\mathcal{U}(2458653.5,2458654.5)$ & $2459756.40002_{-0.00032}^{+0.00029}$ \\ 
$P_b$, ($\mathrm{d}$) & $\mathcal{U}(3.8,4.2)$ & $3.9652482_{-0.0000015}^{+0.0000014}$  \\ 
$q_{1; \mathrm{TESS}}$ &  $\mathcal{N}(0.327,0.05) $  & $0.328\pm0.048$ \\ 
$q_{2; \mathrm{TESS}}$ &  $\mathcal{N}(0.290,0.05) $  & $0.290\pm0.049$ \\ 
$q_{1;\rm  Sloan-z'}$ &  $\mathcal{N}(0.185,0.05) $  & $0.188\pm0.046$ \\ 
$q_{2;\rm  Sloan-z'}$ &  $\mathcal{N}(0.281,0.05) $  & $0.282\pm0.049$ \\ 
$q_{1;\rm  Sloan-i'}$ &  $\mathcal{N}(0.391,0.05) $  & $0.415\pm0.048$ \\ 
$q_{2;\rm  Sloan-i'}$ &  $\mathcal{N}(0.302,0.05) $  & $0.317\pm0.048$ \\ 
$q_{1;\rm  Sloan-r'}$ &  $\mathcal{N}(0.688,0.05) $  & $0.690\pm0.049$ \\ 
$q_{2;\rm  Sloan-r'}$ &  $\mathcal{N}(0.368,0.05) $  & $0.371\pm0.048$ \\ 
$q_{1;\rm I+z}$ &  $\mathcal{N}(0.229,0.05) $  & $0.216\pm0.048$ \\ 
$q_{2;\rm I+z}$ &  $\mathcal{N}(0.290,0.05) $  & $0.292\pm0.049$ \\ 
$q_{1;\rm  1.21\mu \rm m}$ &  $\mathcal{N}(0.185,0.05) $  & $0.182\pm0.045$ \\ 
$q_{2;\rm  1.21\mu \rm m}$ &  $\mathcal{N}(0.281,0.05) $  & $0.282\pm0.045$ \\ 
$\mathrm{gp \ln S_0 (ExTrA 2)}$ &           $\mathcal{U}(-12,3)$  &  $1.79_{-0.30}^{+0.34}$  \\
$\mathrm{gp \ln Q (ExTrA 2)}$ &             $\mathcal{U}(-4,-2)$  &  $-3.933_{-0.050}^{+0.11}$    \\
$\mathrm{gp \ln \omega_0 (ExTrA 2)}$ &      $\mathcal{U}(-4,-1)$  &  $-1.052_{-0.075}^{+0.038}$  \\ 
$\mathrm{gp \ln S_0 (ExTrA 3)}$ &           $\mathcal{U}(-12,3)$ &   $2.49_{-0.31}^{+0.28}$   \\
$\mathrm{gp \ln Q (ExTrA 3)}$ &             $\mathcal{U}(-4,-2)$  &  $-3.947_{-0.039}^{+0.080}$     \\
$\mathrm{gp \ln \omega_0 (ExTrA 3)}$ &      $\mathcal{U}(-4,-1)$  &   $-1.044_{-0.056}^{+0.033}$  \\ 
\hline 
\multicolumn{3}{c}{\textit{Derived parameters}} \\ 
\hline 
$R_\mathrm{b}$ ($\mathrm{R_{\oplus}}$) & \multicolumn{2}{c}{$1.200\pm0.058$}  \\ 
$a_\mathrm{b}/R_\star$ & \multicolumn{2}{c}{$29.3_{-1.6}^{+1.4}$}\\ 
$a_\mathrm{b}$ (AU) & \multicolumn{2}{c}{$0.0272_{-0.0016}^{+0.0015}$  }\\ 
$i_\mathrm{b}$ (deg) & \multicolumn{2}{c}{$89.26_{-0.27}^{+0.33}$ }\\ 
$b_\mathrm{b}$ & \multicolumn{2}{c}{$0.38_{-0.16}^{+0.11}$}  \\ 
$T_\mathrm{tot;b}$ (h) & \multicolumn{2}{c}{$1.020\pm0.017$ }  \\ 
$T_\mathrm{full;b}$ (h) & \multicolumn{2}{c}{$0.896_{-0.021}^{+0.020}$ } \\ 
$T_\mathrm{eq;b}$ (K) &\multicolumn{2}{c}{ $368_{-10}^{+12}$} \\ 
$S_{\rm b}$ ($\rm S_\oplus$) &\multicolumn{2}{c}{ $4.35\pm1.32$} \\
$u_\mathrm{1; TESS}$ & \multicolumn{2}{c}{$0.330\pm0.062$}\\ 
$u_\mathrm{2; TESS}$ & \multicolumn{2}{c}{$0.239_{-0.056}^{+0.061}$}\\ 
$u_\mathrm{1; Sloan-z'}$ & \multicolumn{2}{c}{$0.240_{-0.050}^{+0.054}$ }\\ 
$u_\mathrm{2; Sloan-z'}$ & \multicolumn{2}{c}{$0.185_{-0.048}^{+0.051}$}  \\ 
$u_\mathrm{1; Sloan-i'}$ & \multicolumn{2}{c}{$0.407\pm0.067$} \\ 
$u_\mathrm{2; Sloan-i'}$ & \multicolumn{2}{c}{$0.235\pm0.063$} \\ 
$u_\mathrm{1; Sloan-r'}$ & \multicolumn{2}{c}{$0.617\pm0.082$} \\ 
$u_\mathrm{2; Sloan-r'}$ & \multicolumn{2}{c}{$0.213\pm0.081$}  \\ 
$u_\mathrm{1; I+z}$ & \multicolumn{2}{c}{$0.268_{-0.053}^{+0.056}$} \\ 
$u_\mathrm{2; I+z}$ & \multicolumn{2}{c}{$0.191_{-0.049}^{+0.051}$ }\\
$u_\mathrm{1; 1.21\mu \rm m}$ & \multicolumn{2}{c}{$0.237\pm0.048$} \\ 
$u_\mathrm{2; 1.21\mu \rm m}$ & \multicolumn{2}{c}{$0.184\pm0.047$ }\\
    \end{tabular}}
    \caption{Adopted priors, and fitted and derived parameters from our fit. Uniform priors are indicated as $\mathcal{U}$(lower bound, upper bound), and normal priors are indicated as $\mathcal{N}(\mu, \sigma)$. $\dagger$: {\sc Allesfitter} shifts the epoch to the centre of the data.}
    \label{tab:global_fit}
\end{table}

\section{Upper limit on the planet mass}\label{sec:rvs}

We obtained six spectra of TOI-1080 with NIRPS \citep{Bouchy_2025} spread over three nights: UT 2023-04-14, 2023-04-17, and 2023-04-19. The observations were taken under the NIRPS Guaranteed Time Observation (GTO) program (ID: 111.254T.001), with the initial objective of constraining the mass of TOI-1080\,b. The target was shortly thereafter deemed too challenging, given the small expected semi-amplitude of $\sim$2–3\,m/s and the photon-noise uncertainties of NIRPS RVs for such a relatively faint target, which was still being established at the start of the GTO in April 2023. Here, we use these spectra as a reconnaissance radial velocity (RV) of the star to put an upper limit on the mass of the transiting object.  These RV data discard the scenario that the transit signal is caused by stellar or brown-dwarf companion orbiting the host. 

NIRPS is a high-precision, fiber-fed near-infrared spectrograph (970--1900\,nm) mounted on the ESO 3.6-m telescope at La Silla Observatory and capable of operating simultaneously with HARPS \citep{Mayor_2003}. Each night, we obtained two consecutive 900\,s NIRPS exposures, while HARPS collected a single 1800\,s integration. Both instruments were set to their high-efficiency modes to maximize throughput, the HE mode for NIRPS ($R{\approx}75\,000$, $0\farcs9$ fiber) and the EGGS mode for HARPS ($R{\approx}80\,000$, $1\farcs4$ fiber).  The NIRPS spectra have a SNR of about 30 per pixel in the middle of the $H$ band and the co-added spectrum has a peak SNR of about 60. For comparison, the reddest HARPS order near 690\,nm reached an SNR per pixel of about 5. Although, these data are sufficient for RVs, they cannot be used for accurate stellar characterization of the host. 

The NIRPS data were reduced with the APERO pipeline \citep{Cook_2022}, which delivers telluric-corrected spectra. The HARPS observations were processed using version 3.5 of the original HARPS data reduction software \citep{Lovis_2007}. The radial velocities were extracted with the line-by-line code (\textsc{LBL}, v0.65.012) described in \cite{Artigau_2022}. To improve RV precision, we used high-SNR templates of GJ~643 (M4V) for NIRPS and GJ~3090 (M2V) for HARPS, both observed as part of the NIRPS-GTO.

\begin{figure}
    \includegraphics[width=1\linewidth]{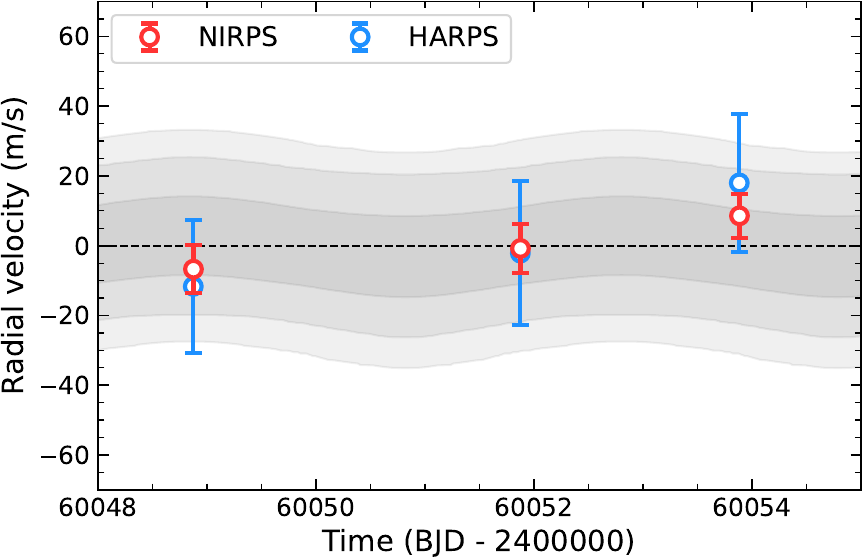}
    \caption{Reconnaissance radial velocity of TOI-1080 with simultaneous NIRPS and HARPS observations. The gray regions show 1-, 2-, and 3-sigma confidence envelopes for a fixed ephemeris circular, Keplerian fit with per-instrument offsets (3 free parameters). We derive a 3-$\sigma$ upper limit of 10.7\,M$_{\oplus}$ for the mass of TOI-1080\,b. \label{fig:nirps_harps}}
\end{figure}

We obtain RV precisions of 7\,m/s with NIRPS (nightly bin) and 20\,m/s with HARPS. We show the resulting RVs in Figure~\ref{fig:nirps_harps}, from which we can derive a 3-$\sigma$ upper limit of 10.7\,M$_{\oplus}$ for the mass of TOI-1080\,b. This upper limit was obtained by fitting a sinusoid with the period and phase fixed to the transit solution (Sect.~\ref{sec:lcmodels}) and allowing separate instrumental RV offsets (3 degrees of freedom).

\section{Discussion}\label{sec:discussion}

TOI-1080 is a nearby ($\sim$25~pc) mid-type M dwarf which hosts at least one transiting planet, with a radius of $R_{\rm p}=1.200\pm0.058~\rm R_\oplus$ and an equilibrium temperature of $T_{\rm eq}=368_{-10.}^{+12}$~K. The planet receives an instellation flux of $S_{\rm b}=4.35\pm1.32~\rm S_\oplus$, placing it inside the `temperate zone', defined by  \citet{Scott2026}
as $0.1~{\rm S_{\oplus}} \leq S_{\rm p}\leq5~\rm S_{\oplus}$ and $T_{\rm eq}\leq400$~K. 

\begin{figure}
    \centering
    \includegraphics[width=\columnwidth]{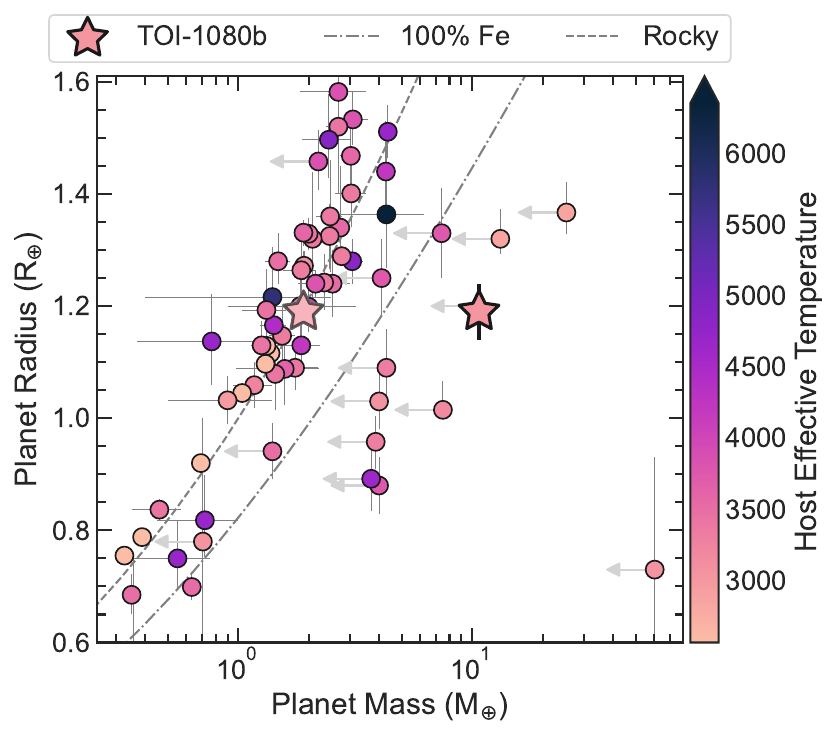}
    \caption{Planets within 50 parsecs with $R_p<1.6~\rm R_\oplus$ from the NASA Exoplanet Archive. Points are coloured according to host star effective temperature and upper limits on masses are indicated with grey arrows. Composition curves for `rocky' and `100\% Fe' are taken from \protect\cite{Zeng2019}. TOI-1080~b is indicted as a star; the mass value plotted is the upper limit we derive in Section \ref{sec:rvs}; we show as a fainter star the mass derived from the mass-radius relations of \protect\cite{Chen2017} using {\sc Forecaster}. }
    \label{fig:mr}
\end{figure}

At present we do not have a mass measurement for TOI-1080~b, only an upper limit. However, using the mass-radius relations presented in \cite{Chen2017} we calculate an estimated mass of $M_{\rm p}=1.75_{-0.57}^{+1.25}~\rm M_\oplus$. We also used {\sc Spright} \citep{Parviainen2024} to check this value and find it to be consistent. Planets with radii $<1.6~\rm R_\oplus$ are likely to be rocky \citep{Rogers2015,Cloutier2020, Parviainen2024} and are thus exciting candidates for atmospheric investigations. Those orbiting bright enough stars are particularly amenable. In Fig.~\ref{fig:mr} we present a mass-radius diagram placing TOI-1080~b in the context of other small, rocky planets within 50~pc of the sun. 

In the remainder of this section we discuss prospects for further analysis of this system. We begin by examining whether precise mass and atmospheric constraints are possible, followed by our insights into the detectability of planetary siblings.

\subsection{Prospects for a precise mass measurement of TOI-1080~b}

Based on the the mass estimate calculated above and the stellar mass derived in Section \ref{sec:star}, we estimate that TOI-1080~b would produce a radial velocity semi-amplitude of $\sim 2.5\rm~m~s^{-1}$. 
With NIRPS, the typical precision achieved was $\sim 7\rm~m~s^{-1}$ with 1800~s integrations. Assuming only RV photon-noise without accounting for stellar activity or additional unknown companions, we calculate that it would have taken at least 200 measurements to obtain a $5\sigma$ mass measurement. This reduces to 70 spectra to measure the planet mass to $3\sigma$. Therefore, the  optimistic estimate of NIRPS telescope time needed would be 100~hours or 36~hours respectively. 

We used the ESPRESSO ETC (exposure time calculator) to estimate the radial velocity precision that could be achieved with this instrument. Given the larger aperture of the VLT, ESPRESSO is well suited to faint stars like TOI-1080. We find that in average observing conditions (airmass $\leq 1.5$ and seeing $\leq1.3\arcsec$) we could achieve radial velocity precisions of $5.6\rm~m~s^{-1}$ with 1800~s exposures. With this setup and not accounting for stellar activity or additional companions, we calculate that 133 or 48 spectra would be needed for 5 or $3\sigma$ mass measurements respectively. Including overheads, this amounts to 85~hours or 30~hours of telescope time, which is a realistic programme for a nearby rocky planet suitable for atmospheric characterisation (see Section \ref{sec:atmos}). Especially given the scarcity of nearby rocky planets amenable to atmospheric characterisation.

There is a strong precedent for precise mass measurements of small rocky planets orbiting low-mass stars with ESPRESSO. For instance, using 44 ESPRESSO spectra, \cite{Lacedelli2024} measured the masses of both planets in the TOI-406 system, consisting of $1.2~\rm R_\oplus$ and $2.2~\rm R_\oplus$ planets with orbital periods of $3.3~\rm days$ and $13~\rm days$ respectively. TOI-406 is an M3 host, and about 2 magnitudes brighter than TOI-1080, so naturally we expect more spectra would be needed in our case. In the case that there are additional planets in the system more telescope time would also be needed to resolve both orbits.


\subsection{Is TOI-1080~b suitable for atmospheric investigations?}
\label{sec:atmos}

\begin{figure}
    \centering
    \includegraphics[width=\columnwidth]{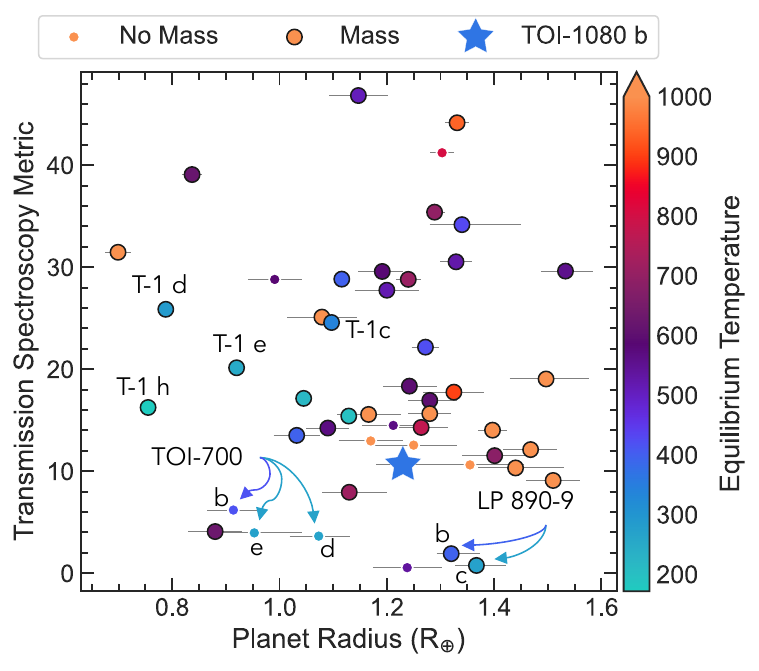}
    \caption{Planets from NASA exoplanet archive with R$<1.6~\rm R_\oplus$ that have already been selected for observations with \textit{JWST} according to TrExoLiSTS \protect\citep[Transiting Exoplanets List of Space Telescope Spectroscopy,][]{Nikolov2022}. Points are coloured according to planetary equilibrium temperature; small circles have no mass measurement and a mass-radius relation is used to calculate their mass. Larger circles have a mass measurement via radial velocities or transit timing variations. }
    \label{fig:TSM}
\end{figure}

There are several methods used to quantify how suitable a planet is for atmospheric characterisation, the most common being the transmission spectroscopy metric \citep[TSM,][]{Kempton2018}. We find that TOI-1080~b has a TSM of 10.6, and we contextualise this number in Fig.~\ref{fig:TSM}. All planets in this figure have $R_{\rm p}<1.6~\rm R_\oplus$ and have already been selected for atmospheric characterisation with \textit{JWST}, according to TrExoLiSTS \citep{Nikolov2022}. TOI-1080~b has a higher TSM than TOI-700~b,d and e, \citep{Gilbert2020,Gilbert2023} and both planets in the LP~890-9 \citep{Delrez2022} system. We also calculate the ESM \citep[emission spectroscopy metric,][]{Kempton2018} for TOI-1080~b and find a value of 1.4, comparable to TRAPPIST-1~c which has been observed in emission with \textit{JWST}'s MIRI and had a thick atmosphere ruled out \citep{Gillon2025}.

Another way to select suitable targets for atmospheric characterisation is to attempt to determine a priori their likelihood of possessing an atmosphere in the first place. For instance, the `cosmic shoreline' \citep{Zahnle2017}, which is a theoretical boundary between planets with and without atmospheres, based on instellation flux received and planetary escape velocity. This framework was recently updated by \cite{Ji2025}, and they use their parametric definition of the CO$_2$ cosmic shoreline to calculate a priority score for rocky exoplanets. 
We use Equation 7 of this paper to calculate the CO$_2$ cosmic shoreline for TOI-1080~b, defined as the instellation at which it has 90\% probability of retaining a CO$_2$ atmosphere. We find a value of $11.04~\rm S_\oplus$, placing TOI-1080~b comfortably below this boundary with its instellation of $S_{\rm b}=4.35\pm1.32~\rm S_\oplus$. This gives a priority score of 0.40; for context, this would place TOI-1080~b 28$^{\rm th}$ on the list of top 50 rocky planets for atmospheric characterisation, had it been known at the time \citet{Ji2025} was published. 

The habitable zone, defined as the region around a star where a rocky planet with atmosphere may retain surface liquid water, is located between 0.047 and 0.124 au for TOI-1080 \citep{Kopparapuetal2013}. TOI-1080~b is located outside the habitable zone at 0.0272 $\pm$ 0.0016 au. Given that TOI-1080~b receives an instellation flux $> 1.48 S_\oplus$, it is expected to start a runway greenhouse, that may result in a massive oxygen (O$_2$) atmosphere (from a few hundreds to a few thousands bars) after water molecules are photolyzed by UV radiation and hydrogen escapes from the planet \citep{LugerBarnes2015,Schaeferetal2016,KiteSchaefer2021,Cherubimetal2025}. Rich abiotic O$_2$ atmospheres are a robust prediction for short period planets around M dwarfs, like TOI-1080~b, and their characterization could be considered as an intermediate and necessary step towards the search of spectroscopic biosignatures \citep{Cherubimetal2025}.

Thus far, no robust detections of a secondary atmosphere have been made for temperate rocky exoplanets observed in transmission with \textit{JWST} \citep[e.g.][]{Allen2025,Bennett2025,Luque2025}. \cite{Kreidberg2025} show that in most cases, a 5 scale height atmosphere would still be hidden in the noise for most planets observed, despite the outstanding precision of \textit{JWST} and the large amount of time already invested. Therefore the Rocky Worlds DDT Programme \citep{Redfield2024,Xue2025} has selected nine of the most exciting planets to observe in emission, as with this approach a thick atmosphere can be ruled out if there is no evidence of heat redistribution on tidally locked planets. TOI-1080~b is an excellent target for this programme; its priority  score as defined by \cite{Ji2025} is higher than four out of nine targets selected, and its ESM is higher than two of the selected targets.

\subsection{Additional transiting planets in the TOI-1080 system}\label{sec:injection}

We first assessed the detection limits of additional transiting planets in TOI-1080 by running  Transit Least Squares  \citep[{\sc TLS}]{hippke2019} on the residuals of \textit{TESS} light curve to the best fit model described in  Section~\ref{sec:lcmodels}. 
We do not find any significant signal. We then did an injection-recovery test on the same \textit{TESS} residuals where we used \textsc{Batman} \citep{kreidberg2015} to generate synthetic transit signals of planets with radii between  0.38 (like Mercury) to 5 R$_\oplus$ and with orbital periods between 0.5 to 30 days. In all the cases, circular orbits and 90 degrees of inclination were assumed while the mid-transit times were randomly selected between the minimum and maximum time in the  \textit{TESS} light curve. Once the planetary signal was injected in the residuals, we used {\sc TLS} to search for the transits. If the period recovered was within 5$\%$ of the injected period, we considered a positive detection. 

\begin{figure}
    \includegraphics[width=1.0\columnwidth,trim=0.4cm 0.44cm 0.395cm 0.44cm, clip]{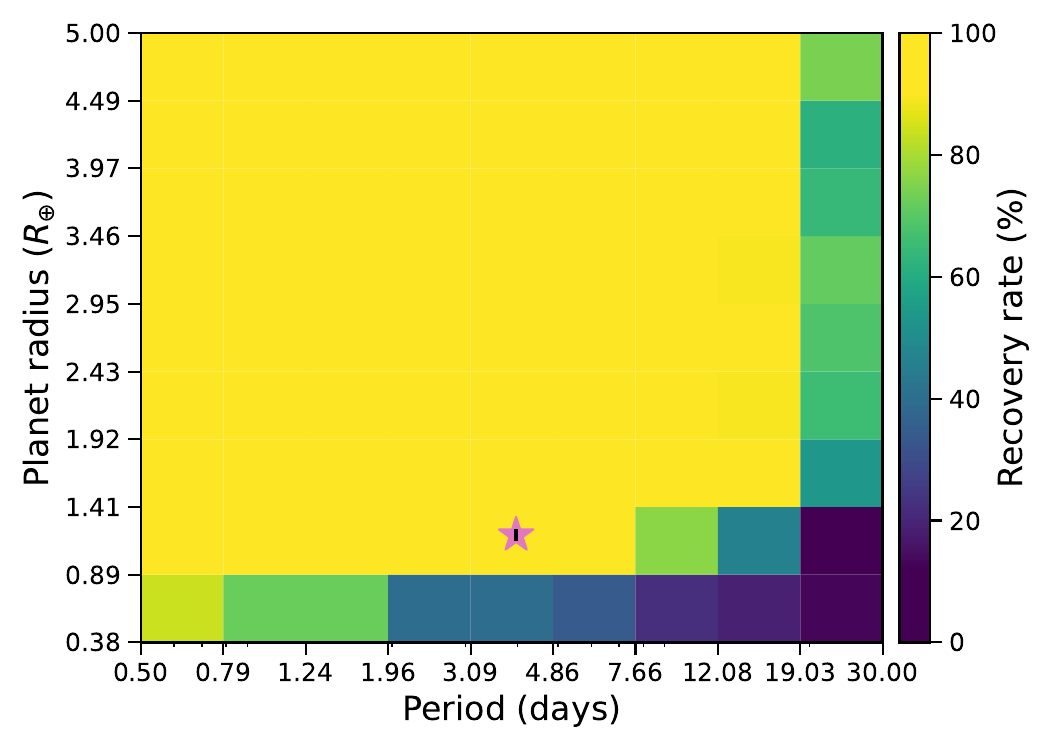}
    \caption{Injection-recovery test performed on the residuals of the \textit{TESS} light curve of TOI-1080. High and low recovery rates are indicated with light and dark colors, respectively.  TOI-1080~b is marked by the pink star and the errors on the planet radius are marked by the black vertical line. The errors on its orbital period are too small to be seen in the plot. 
  These injection-recovery tests are able to discard other planets in the system 
     }
         \label{fig:injection}
\end{figure}

We can rule out the existence of transiting planets down to $\sim$0.9 R$_\oplus$ with orbital periods shorter than $\sim$7.7 days in the TOI-1080 system, as in our simulations the planets were recovered in this period range in $\gtrsim$90\% of the injections. 
We are also able to rule out larger planets than  1.4 R$_\oplus$  with the \textit{TESS} data with periods up to 19 days (better than 90\% recovery of the injections). Because the three \textit{TESS} sectors are separated by 2 years or more, the detection rate starts quickly declining beyond 19-day orbital periods for all planet radii, as expected for individual 28-d long \textit{TESS} sectors. 
However, planet multiplicity is common for M-dwarf hosts. 
TOI-1080 is not planned to be observed with \textit{PLATO} based on current LOPS2 field definition, so revealing further planets will need dedicated ground-based monitoring campaign, more \textit{TESS} sectors or alternatively an intensive RV campaign.

Using all the available SPECULOOS 
observations (243 hours), we computed the percentage of phase coverage for all orbital periods from 0.1 to 15 days, following a similar approach to that described in \cite{Sebastian2021}. We obtained an effective coverage of 80\% for $P_{\rm max}=12.8$ days. This effective coverage metric is defined as the integral of the percentage of phase coverage over the period range from $P=0.1$ days to $P_{\rm max}=12.8$ days.

\begin{figure}
    \centering
    \includegraphics[width=\columnwidth,trim=1.4cm 0.4cm 2.5cm 1.3cm,clip]{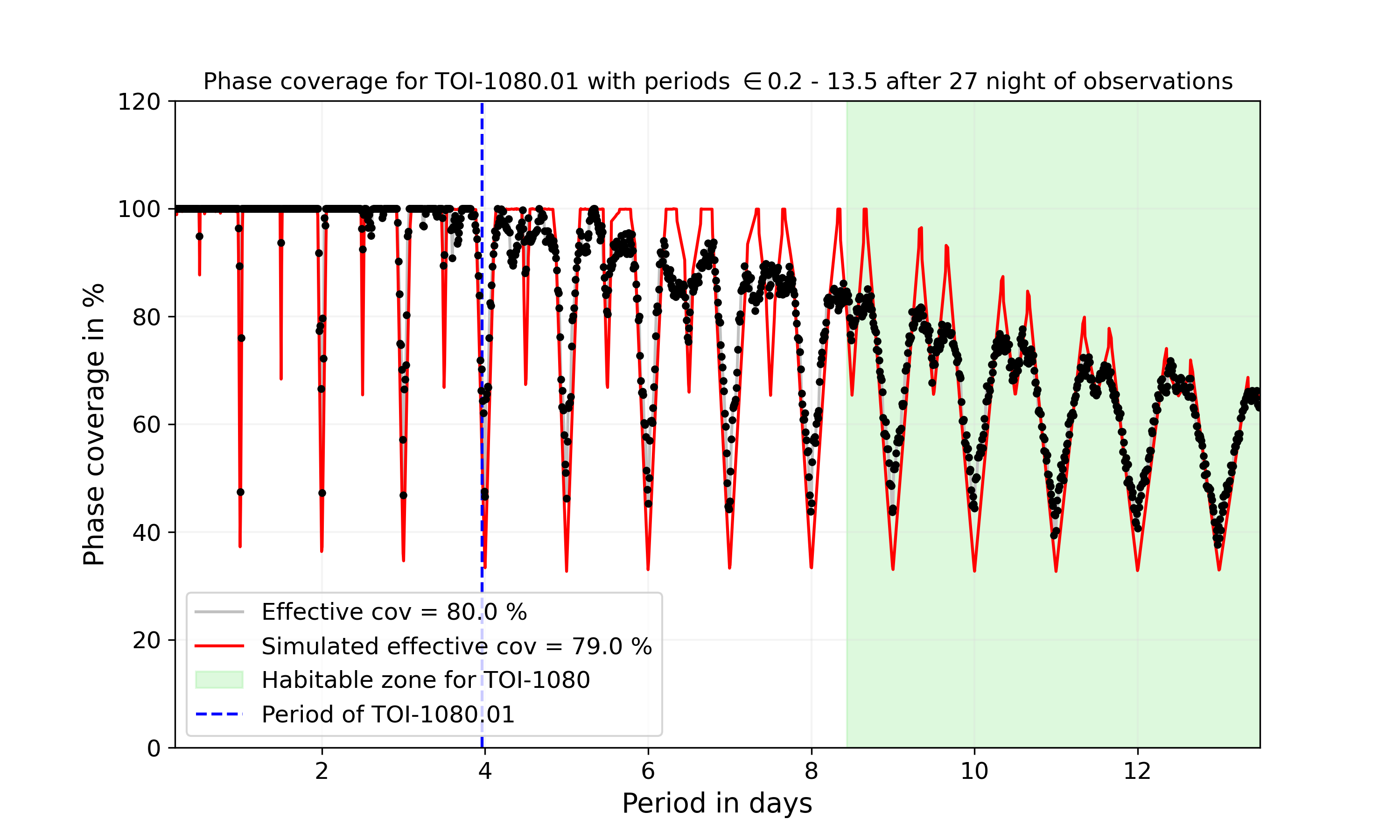}
    \caption{Phase coverage as a function of the orbital period of an hypothetical additional planet around TOI-1080, observed for 243 hours in total by SPECULOOS.}
    \label{fig:phase_coverage_obs_vs_simu}
\end{figure}

Our goal is to estimate the number of additional nights required to achieve an effective phase coverage of 80\% over the full optimistic habitable zone (HZ) of TOI-1080, thereby allowing us to significantly detect or rule out HZ planets in this system. To this end, we simulated 27 nights of observations—matching the number of nights obtained with SPECULOOS—each lasting 8 hours. We verified that the resulting percentage of phase coverage and effective coverage from the simulations are consistent with those derived from the real observations. Figure \ref{fig:phase_coverage_obs_vs_simu} confirms this agreement, showing that the simulated and observed coverages are indeed very similar.

\begin{figure}
    \centering
    \includegraphics[width=\columnwidth,trim=1.4cm 0.4cm 2.5cm 1.3cm,clip]{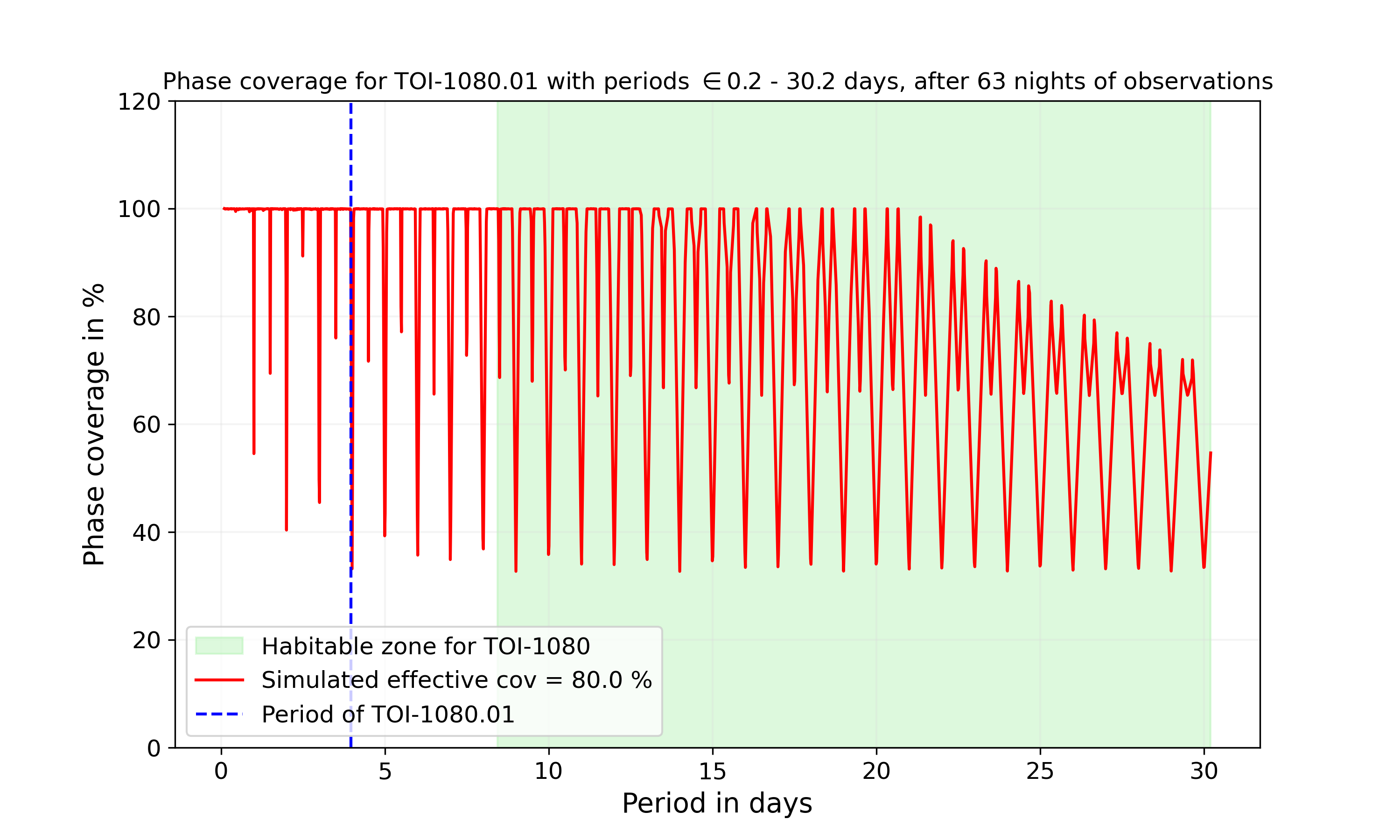}
    \caption{Simulation of the phase coverage as a function of the orbital period for 63 nights to reach effective coverage of 80\% for $P_{max}$ set to the period of the outer bound of optimist habitable zone. }
    \label{fig:simulated_phase_coverage_HZ}
\end{figure}

We then extended the simulations to N nights, where $N$ is the number of nights required for the effective coverage at $P_{\rm max}$ = Period of the outer HZ limit (computed using \cite{Kopparapu2014}) to reach 80\%. We find $N$=67 nights, implying that 36 additional nights (63 – 27) of monitoring are needed to fully explore the habitable zone of TOI-1080 (see Figure \ref{fig:simulated_phase_coverage_HZ}).

\section{Summary}\label{sec:summary}

In this work we have presented the detection and validation of TOI-1080~b, a temperate $R_{\rm p}=1.200\pm0.058~\rm R_\oplus$, likely rocky planet orbiting an M4 host $\sim25$~pc away. We have shown that this planet is suitable for mass measurements with radial velocity with ESPRESSO, as well as atmospheric characterisation via transmission spectroscopy, and would be a prime target for the \textit{JWST+HST} Rocky Worlds DDT programme. 
Although inner to the TOI-1080 habitable zone, TOI-1080~b could potentially have a CO$_2$ or a massive O$_2$ atmosphere amenable for characterization. 
We rule out additional transiting planets with radii $> 0.9~R_\oplus$ and periods between 0.5 and 7.7~days and for periods up to 19~days we can rule out planets larger than 1.4~R$_\oplus$
in the system based on the \textit{TESS} photometry. Given the scarcity of nearby terrestrial planets suitable for in-depth characterisation, TOI-1080~b represents an exciting addition to the known sample. 


\section*{List of Affiliations}
$^{1}$ \mex \\
$^{2}$ \oxfordastro \\
$^{3}$ \oxfordmag \\
$^{4}$ \birmingham \\
$^{5}$ \iac \\
$^{6}$ \liege \\
$^{7}$ \miteaps \\
$^{8}$ \geneva \\
$^{9}$ \trottier \\
$^{10}$ \mariecurie \\
$^{11}$ \aim \\
$^{12}$ \mitkavli \\
$^{13}$ \sandiego \\
$^{14}$ \icn \\
$^{15}$ \vandy \\
$^{16}$ \sfasu \\
$^{17}$ \grenoble \\
$^{18}$ \bern \\
$^{19}$ \nasaames \\
$^{20}$ \oac \\
$^{21}$ \conicet \\
$^{22}$ \uname \\
$^{23}$ \cfa \\
$^{24}$ \pest \\
$^{25}$ \sharjah \\
$^{26}$ \cadi \\
$^{27}$ \cambridge \\
$^{28}$ \liegestar \\
$^{29}$ \iaa \\
$^{30}$ \uniturkey \\
$^{31}$ \obsturkey \\
$^{32}$ \riogrande \\
$^{33}$ \ull \\
$^{34}$ \instiporto \\
$^{35}$ \uniporto \\

\section*{Acknowledgements}
The authors of this paper would like to thank Rob Simcoe at MIT for acquiring the FIRE spectrum.
YGMC, AK, and MPM are partially supported by UNAM PAPIIT-IG101224 and by the Swiss National Science Foundation IZSTZ0\_216537. 
GD acknowledges funding from Magdalen College, Oxford.
Funding for the \textit{TESS} mission is provided by NASA's Science Mission Directorate. We acknowledge the use of public \textit{TESS} data from pipelines at the \textit{TESS} Science Office and at the \textit{TESS} Science Processing Operations Center. Resources supporting this work were provided by the NASA High-End Computing (HEC) Program through the NASA Advanced Supercomputing (NAS) Division at Ames Research Center for the production of the SPOC data products.
This research has made use of the Exoplanet Follow-up Observation Program (ExoFOP; DOI: 10.26134/ExoFOP5) website, which is operated by the California Institute of Technology, under contract with the National Aeronautics and Space Administration under the Exoplanet Exploration Program. This paper includes data collected by the \textit{TESS} mission that are publicly available from the Mikulski Archive for Space Telescopes (MAST).
Data obtained at the ESO la Silla Observatory in Chile.
The research leading to these results has received funding from  the ARC grant for Concerted Research Actions, financed by the Wallonia-Brussels Federation. TRAPPIST is funded by the Belgian Fund for Scientific Research (Fond National de la Recherche Scientifique, FNRS) under the grant PDR T.0120.21.
MG and EJ are F.R.S.-FNRS Research Directors.
This research has made use of the NASA Exoplanet Archive, which is operated by the California Institute of Technology, under contract with the National Aeronautics and Space Administration under the Exoplanet Exploration Program.
This research has made use of data provided by the portal exoplanet.eu of The Extrasolar Planets Encyclopaedia.
This work makes use of observations from the LCOGT network. Part of the LCOGT telescope time was granted by NOIRLab through the Mid-Scale Innovations Program (MSIP). MSIP is funded by NSF.
The ULiege's contribution to SPECULOOS has received funding from the European Research Council under the European Union's Seventh Framework Programme (FP/2007-2013) (grant Agreement n$^\circ$ 336480/SPECULOOS), from the Balzan Prize and Francqui Foundations, from the Belgian Scientific Research Foundation (F.R.S.-FNRS; grant n$^\circ$ T.0109.20), from the University of Liege, and from the ARC grant for Concerted Research Actions financed by the Wallonia-Brussels Federation. 
The Cambridge contribution is supported by a grant from the Simons Foundation (PI Queloz, grant number 327127).
The Bern contribution is supported by the Swiss National Science Foundation (PP00P2-163967, PP00P2-190080 and the National Centre for Competence in Research PlanetS). The Birmingham contribution to SPECULOOS has received fund from the European Research Council (ERC)
under the European Union's Horizon 2020 research and innovation programme (grant agreement n$^\circ$ 803193/BEBOP), from the MERAC foundation, from the Science and Technology Facilities Council (STFC; grants n$^\circ$ ST/S00193X/1, ST/W000385/1 and ST/Y001710/1) and from the ERC/UKRI Frontier Research Guarantee programme (EP/Z000327/1/CandY).
We acknowledge funding from the European Research Council under the ERC Grant Agreement n. 337591-ExTrA.
Visiting Astronomer at the Infrared Telescope Facility, which is operated by the University of Hawaii under contract 80HQTR24DA010 with the National Aeronautics and Space Administration.
This paper includes data gathered with the 6.5 meter Magellan Telescopes located at Las Campanas Observatory, Chile.
We thank the Swiss National Science Foundation (SNSF) and the Geneva University for their continuous support of the NIRPS planet search programmes. This work has been in particular carried out in the frame of the National Centre for Competence in Research PlanetS supported by the SNSF under grant 51NF40\_205606.
Funding for KB was provided by the European Union (ERC AdG SUBSTELLAR, GA 101054354).
This material is based upon work supported by the National Aeronautics and Space Administration under Agreement No.\ 80NSSC21K0593 for the program ``Alien Earths''.
The results reported herein benefited from collaborations and/or information exchange within NASA’s Nexus for Exoplanet System Science (NExSS) research coordination network sponsored by NASA’s Science Mission Directorate.
This material is based upon work supported by the European Research Council (ERC) Synergy Grant under the European Union’s Horizon 2020 research and innovation program (grant No.\ 101118581---project REVEAL).
EJ acknowledges funding from CONICET, under project PIBAA-CONICET ID-73669.
RP acknowledge funding from CONICET under project PIBAA-CONICET ID-73811. 
SP acknowledges the support from UNAM-PAPIIT-IG101224 and from SECIHTI through the Becas Nacionales de Posgrado program. 
Author F.J.P acknowledges financial support from the Severo Ochoa grant CEX2021-001131-S funded by MCIN/AEI/10.13039/501100011033 and Ministerio de Ciencia e Innovación through the project PID2022-137241NB-C43. 
JIGH acknowledges financial support from the Spanish Ministry of Science, Innovation and Universities (MICIU) project PID2023-149982NB-I00.
The research activities of the Board of Observational and Instrumental Astronomy (NAOS) at the Federal University of Rio Grande do Norte are supported by continuous grants from the Brazilian funding agency Conselho Nacional de Desenvolvimento Cient\'ifico (CNPq). J.R.M.  acknowledges a CNPq research fellowship (grant No. 308928/2019-9). 
 NCS is co-funded by the European Union (ERC, FIERCE, 101052347). Views and opinions expressed are however those of the author(s) only and do not necessarily reflect those of the European Union or the European Research Council. Neither the European Union nor the granting authority can be held responsible for them. This work was supported by FCT - Fundação para a Ciência e a Tecnologia through national funds by these grants: UIDB/04434/2020 DOI: 10.54499/UIDB/04434/2020, UIDP/04434/2020 DOI: 10.54499/UIDP/04434/2020.

\section*{Data Availability}

\textit{TESS} data products are available via the MAST portal at \url{https://mast.stsci.edu/portal/Mashup/Clients/Mast/Portal.html}. Follow-up photometry and high resolution imaging data for TOI-1080 are available on ExoFOP at \url{https://exofop.ipac.caltech.edu/tess/target.php?id=161032923}. These data are freely accessible to ExoFOP members immediately and are publicly available following a one-year proprietary period. NIRPS and HARPS spectra are available on the ESO Portal following a 1 year proprietary period.



\bibliographystyle{mnras}
\bibliography{1080} 




\appendix


it can be placed in an Appendix which appears after the list of references.


\bsp	
\label{lastpage}
\end{document}